\documentclass[aps,a4paper,10pt,twocolumn,nofootinbib]{revtex4} % eqsecnum
\usepackage[T1]{fontenc}
\usepackage{pslatex}
\usepackage{datetime}
\usepackage{amsmath}
\usepackage{amsfonts}
\usepackage{amsfonts}
\usepackage{mathrsfs}
\usepackage[mathscr]{euscript}
\usepackage[dvips]{graphicx}
\usepackage{fancyhdr}
\usepackage{colordvi}
\usepackage{hyperref} % [hypertex=true]
\usepackage{color}
\usepackage{bm}
\usepackage{soul}

\def\overstrike#1#2{{\setbox0\hbox{$#2$}\hbox to \wd0{\hss
    $#1$\hss}\kern-\wd0\box0}}
\newcommand{\convol}{\star}
\newcommand{\RealPart}{\mathbb{R}\textrm{e}}
\newcommand{\ImagPart}{\mathbb{I}\textrm{m}}

\renewcommand{\Vec}{\bm}

        \DeclareMathOperator{\sgn}{sgn}

        \DeclareMathOperator{\arctanh}{arctanh}

        \DeclareMathOperator{\cross}{\times}
        \DeclareMathOperator{\grad}{\nabla}

\newdateformat{yymmdddate}{\THEYEAR/\twodigit{\THEMONTH}/\twodigit{\THEDAY}}

\begin{document}
\title{How to be causal: time, spacetime, and spectra}
\author{Paul Kinsler}
\email{Dr.Paul.Kinsler@physics.org}
\affiliation{
  Blackett Laboratory, Imperial College,
  Prince Consort Road,
  London SW7 2AZ, 
  United Kingdom.
}

\lhead{\includegraphics[height=5mm,angle=0]{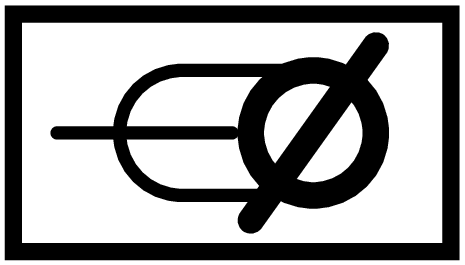}~~HOW2C}
\chead{How to be causal}
\rhead{
\href{mailto:Dr.Paul.Kinsler@physics.org}{Dr.Paul.Kinsler@physics.org}\\
\href{http://www.kinsler.org/physics/}{http://www.kinsler.org/physics/}
}
%\lfoot{\thesection . \thesubsection; ~~~~ (\yymmdddate\today:\currenttime) 
% \href{http://localhost/}{(0)}}
%\rfoot{{\large \emph{Not for redistribution}}}

\begin{abstract}

I explain a simple definition of causality in widespread use, 
 and indicate how it links to the Kramers Kronig relations.
The specification of causality 
 in terms of temporal differential eqations 
 then shows us the way to write down dynamical models 
 so that their causal nature \emph{in the sense used here}
 should be obvious to all.
To extend existing treatments of causality
 that work only in the frequency domain,
 I derive a reformulation of the long-standing Kramers Kronig relations 
 applicable not only to just temporal causality, 
 but also to spacetime ``light-cone'' causality 
 based on signals carried by waves.
I also apply this causal reasoning to Maxwell's equations, 
 which is an instructive example since their casual properties
 are sometimes debated.

\end{abstract}

\date{\today}
\maketitle
\thispagestyle{fancy}

%
%\tableofcontents

%
% =======================================================================
\section{Introduction}\label{S-intro}

Causality\footnote{\noindent{
  \scriptsize\emph{Published in Eur. J. Phys. \textbf{32}, 1687 (2011),
  subsequently updated here.}\\
% {\url{http://iopscience.iop.org/0143-0807/32/6/022}}\\
 \tiny{\textbf{Statement required by the publisher of the EJP:}
 This is an author-created, un-copyedited version of an article
  accepted for publication in the European Journal of Physics. 
 IOP Publishing Ltd is not responsible for any errors or omissions
  in this version of the manuscript or any version derived from it. 
 The definitive publisher-authenticated version is available online at 
 \href{http://iopscience.iop.org/0143-0807/32/6/022}{doi:10.1088/0143-0807/32/6/022}.}
}}
 is a basic concept in physics - 
 so basic, 
 in fact, 
 that it is hard to conceive of a useful model in which effects 
 do not have causes.
Indeed, 
 the whole point of a physical model could 
 be said to describe the process of cause and effect
 in some particular situation.
But what do we generally mean by word like ``causality'', 
 and phrases such as ``cause and effect''?
Usually, 
 we mean that the cause of any event 
 must not be later than any of its effects.
But even such simple-sounding statements
 are rarely as uncomplicated as they seem:
 when trying to clarify the details and built-in assumptions, 
 it is easily possible to get into philosophical discussions
 \cite{Lobo-2007,Lucas-Causality}, 
 issues regarding statistical inference and induction \cite{Pearl-Causality}, 
 or particular physical arguments \cite{Rolnick-Causality}.
Here I instead follow the physics tradition
 characterized
 by Mermin as ``shut up and calculate'' \cite{Mermin-1989pt-shutup}.
But \emph{what} should we calculate, 
 and \emph{how}?

Note that common expressions
 such as $F=ma$ do not express a causal relationship
 in the sense used here.
They provide no means of telling
 whether the force $F$ is supposed to cause an acceleration $a$, 
 or $a$ cause $F$, 
 or even if the equation is instead intended as a constraint of some kind.
Instead, 
 we start with differential equations containing temporal derivatives, 
 which are \emph{open-ended specifications for the future behaviour}.
They require only a knowledge of initial conditions
 and the on-going behaviour of the environment to solve.

Additionally, 
 since any effect $R$ must not occur before its cause $Q$, 
 a mathematical expression for $R$
 must become non-zero only after that cause. 
This desired behaviour matches that of the
 mathematical step function $h(t)$,
 which has $h(t)=0$ for $t < 0$, 
 and $h(t)=1$ for $t \ge 0$.
The definition of causality applied by this temporal step function
 is the same as that enforced 
 in the frequency domain by the famous 
 Kramers Kronig relations \cite{Bohren-2010ejp,LandauLifshitz}.
Therefore I call the time-step causality discussed here 
 ``KK causality'', 
 to distinguish it from alternative definitions.

After describing the role of differential equations
 and how they can generate the time-step criteria
 in section \ref{S-causalDE}, 
 I consider causality in the spectral domain in section \ref{S-KK}, 
 and discuss typical models in section \ref{S-Pendulumn}.
Causality and Maxwell's Equations is discussed in section \ref{S-Maxwell},
 followed by the development of a new spacetime formulation
 of the Kramers Kronig relations in section \ref{S-waves}.
After a discussion in section \ref{S-simul},
 I summarize in section \ref{S-summary}.
There are also two appendices (A and B) --
 not present in the published version --
 %(\ref{S-appdx-discrete}, \ref{S-appdx-types})
 which consider the results of section \ref{S-causalDE}
 applied to discretized (or numerical) models, 
 and then the role of spacetime transformations when
 attributing of causes and effects.

This undergraduate level presentation 
 introduces causality as a topic in itself,
 making
 specific reference to the construction of causal models, 
 rather than being a brief diversion on the way to discussing
 the Kramers Kronig relations, 
 as is the case in many textbooks.
It details the connection between time and frequency domain representations
 and then extends this to a full spacetime ``wave-cone'' causality.
I envisage that this discussion could be incorporated 
 most easily into relativity or electromagnetism courses, 
 although parts could be integrated into courses on 
 mechanics or wave motion.

%
% =======================================================================
\section{Causal differential equations}\label{S-causalDE}

\begin{figure}
\includegraphics[angle=-0,width=0.32\columnwidth]{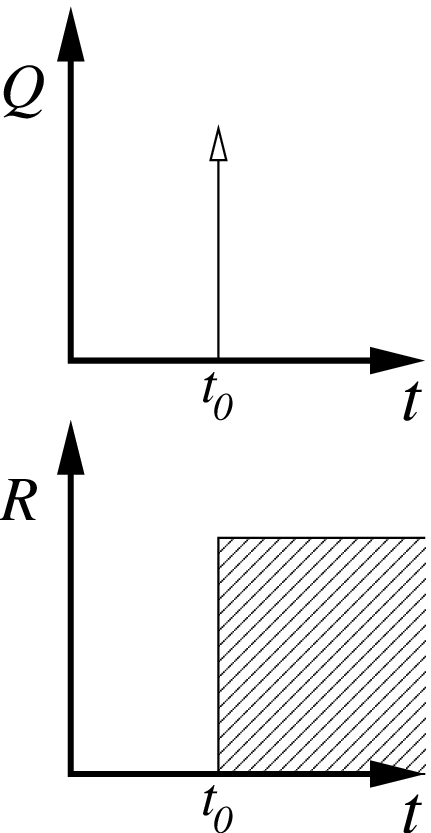}
\caption{
A delta function impulse at a time $t_0$ 
 is the cause $Q=\delta(t-t_0)$ 
 of an effect (a step-change in value) on $R=h(t-t_0)$,
 as a result of the causal model
 specified in eqn. \eqref{eqn-basicRQ}. 
}
\label{fig-RQ}
\end{figure}

Let us first write down a simple model, 
 where some system $R$ responds to its local environment $Q$.
Here $R$ can be any quantity -- 
 e.g. 
 a position or velocity, 
 a level of excitation of some system, 
 and even -- 
 if a position $\Vec{r}$ is also specified --
 a probability distribution or wave function.
Likewise $Q$ might be anything, 
 depending on some pre-set behaviour, 
 the behaviour of $R$ or other systems, 
 or (e.g.) spatial derivatives of fields, 
 potentials, distributions, 
 and so on.
Whatever the specific meaning of $R$ (or indeed of $Q$), 
 we start by writing
 the \emph{simplest possible} differential equation\footnote{See also e.g. 
  \cite{Corduneanu-FECAUSAL,Corduneanu-2000jde} for mathematical details.}. 
~
\begin{align}
  \partial_t R(t)
&=
 Q(t)
,
\label{eqn-basicRQ}
\end{align} 
where $\partial_t$ is just the time derivative $d/dt$.
To determine how causal this model equation is, 
 consider the case where the environment
 contains a \emph{simple possible} cause:
 a brief delta-function impulse, 
 where $Q(t) = Q_0 \delta(t-t_0)$.
Reassuringly,
 if I integrate eqn. \eqref{eqn-basicRQ}, 
 then $R(t)$ will gain a step at $t_0$ --
 i.e. the effect of the impulsive $Q$
 is to cause $R$ to increase discontinuously by $Q_0$ at $t_0$; 
 as depicted on fig. \ref{fig-RQ}.
Thus we see how the step function $h(t)$
 arises directly from the most basic temporal differential equation.
If we apply $h$ as a filter to $R$,
 but it changes nothing 
 (in the example above, $R(t) = R(t) h(t-t_0)$),
 then we know that $R(t)$ is causal, 
 since all effects occur after the cause $Q=Q_0\delta(t-t_0)$.

I describe this situation, 
 where an impulse gives rise to a stepped response
 %(as opposed to e.g. a ramp or some smoother response)
 as ``barely causal'', 
 because part of the effect is simultaneous
 with the cause.
Other responses
 (e.g. see section \ref{S-Pendulumn})
 are usually more complicated and contain higher-order time derivatives,
 leading to a more gradual response.
E.g., 
 a differential equation with 
 second order time derivatives has a ramp-like (linear) response
 to a delta function cause,
 whereas third order derivatives give a quadratic response.

Simple examples from kinematics can illustrate
 the meaning of ``causal'' as used here.
If we were to write down $\partial_t x = v$, 
 then we could make the statement that 
 ``velocity $v$ causes a change in position $x$''; 
 likewise $\partial_t v = a$ means that 
 ``acceleration $a$ causes a change in velocity $v$''; 
 and $\partial_t^2 x = a$ means
 ``acceleration $a$ causes changes in position $x$''.

More general differential equations 
 can be written down 
 as weighted sums of different orders of time derivatives
 e.g. 
~
{\begin{align}
  \sum_{n=0}^{N}
  T_n %\left(T_n\right)^n
  \partial_t^n R(t)
&=
 \sum_{m=0}^{N-1}
  a_m
  \partial_t^m
  Q(t)
,
\label{eqn-multiRQ}
\end{align}}
 for parameters $T_n$, $a_m$, 
 and a defined maximum order of derivatives $N$, 
 with $T_N \neq 0$.
These will remain KK causal as long as the highest
 derivative on the RHS
 is of lower order than that
 on the LHS \cite{Kinsler-2010pra-lfiadc}.

We might consider recasting the differential equations 
 used here in an integral form; 
 e.g. for an evolution starting at a time $t_i$, 
 eqn. \eqref{eqn-basicRQ} becomes
~
\begin{align}
  R(t) = \int_{t_i}^t Q(t') dt'
,
\label{eqn-basicRQ-integral}
\end{align}
 although in most cases
 this is not as easy as writing down the differential equation.
Also, 
 as discussed next, 
 differential equations make it 
 easier to consider the spectral properties.
And on a more intuitive note, 
 writing down a differential equation does not imply you have solved it --
 it is a notation more compatible
 with the concept of an unknown future outcome,
 dependent on as-yet unknown future causes
 \cite{Kinsler-2014arxiv-negfreq}.
This point of view becomes clearer when discretized 
 forms of the differential models are considered, 
 as done in the appendix (\ref{S-appdx-discrete}).
Further, 
 the ambitious may find it interesting to 
 consider the causal set approach \cite{Rideout-S-2000prd}
 to how a universe universe extends itself into its future.

%
% =======================================================================
\section{Causality and spectra}\label{S-KK}

Often, 
 the more complicated a model response is, 
 the more likely it is that its response will be analyzed
 in the frequency domain.
This might be either because an experiment has recorded
 spectral data directly, 
 or time varying data $S(t)$ has been converted
 into a spectrum $\tilde{S}(\omega)$
 using a Fourier transform
 \cite{WfmMathWorld-FourierTransform,Davies-IntegralTfs}.
Although it is common to write down 
 the individual sin and cosine Fourier transforms, 
 it is most convenient to combine them
 using $e^{-\imath \omega t} = \cos(\omega t) + \imath \sin(\omega t)$, 
 giving
~
\begin{align}
  \tilde{S}(\omega)
&=
  \frac{1}{\sqrt{2\pi}}
  \int_{-\infty}^{+\infty}
    S(t)
    e^{-\imath \omega t}
    dt
.
\label{eqn-FourierTf}
\end{align}
Note that even for real-valued $S(t)$, 
 the spectrum $\tilde{S}(\omega)$ can be complex valued.
If $S(t)$ is consistent with casuality, 
 then its spectrum $\tilde{S}(\omega)$ must also,
 and this insistence that measured spectral data 
 must be consistent with causality 
 can be of considerable use \cite{KKinOMR}.
So useful, 
 in fact, 
 that even quite long articles on causality and spectra \cite{Toll-1956pr}
 can get away without any discussion of time-domain dynamics at all!

Let us therefore take our simple eqn. \eqref{eqn-basicRQ} 
 and either Fourier transform it,
 or take the mathematical shortcut of assuming 
  an $\exp(-\imath \omega t)$ time dependence.
Since $\partial_t A(t)$ transforms to $- \imath \omega \tilde{A}(\omega)$, 
 we get
~
\begin{align}
  - \imath \omega \tilde{R}(\omega)
&=
    \tilde{Q}(\omega)
\\
\Longrightarrow \qquad
  \tilde{R}(\omega)
&=
 +
  \imath
  \frac{\tilde{Q}(\omega)}
       {\omega}
.
\label{eqn-RQ-spectra}
\end{align}
If $Q(t) = Q_0 \delta(t)$ is a delta function, 
 then its spectrum is a constant with $\tilde{Q}(\omega) = Q_0$, 
 so that
~
\begin{align}
  \tilde{R}(\omega)
&=
 +
  \imath
  \frac{Q_0}
       {\omega}
.
\label{eqn-RQ-spectra-solve}
\end{align}

Since we already know that the solution for $R$ contains a 
 step at $t_0$, 
 then we now also know (and can check)
 that the Fourier transform of a step function
 is proportional to $1/\omega$.
Returning also to the more general differential form 
 in eqn. \eqref{eqn-multiRQ}, 
 we see that since the LHS has higher order time derivatives 
 than the RHS, 
 any rearrangement to put only $R$ on the LHS 
 (as in eqn. \eqref{eqn-RQ-spectra-solve})
 will result in an RHS that falls off at least as fast as $1/\omega$.

An important and useful way of 
 checking and /or enforcing causality on spectra
 are the 
 Kramers Kronig (KK) relations \cite{Bohren-2010ejp,LandauLifshitz,KKinOMR}.
Although derivations are complicated, 
 their basic construction is based on two concepts:

\begin{figure}
\includegraphics[angle=-0,width=0.72\columnwidth]{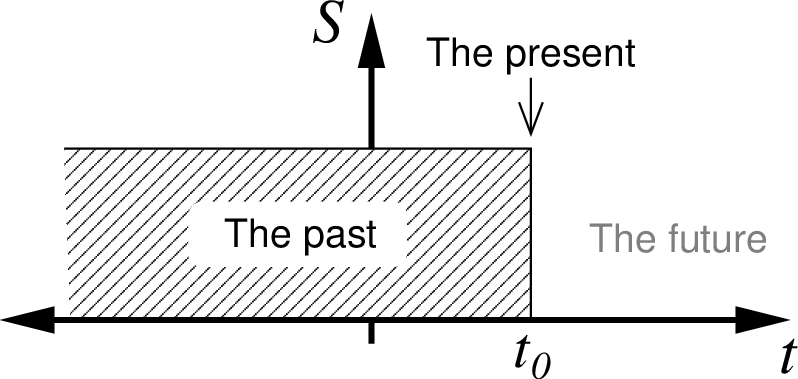}
\caption{
A historical record $S(t)$ taken at a time $t_0$ 
 can only contain data prior to that time.
}
\label{fig-thepast}
\end{figure}

\begin{enumerate}

\item
The Hilbert transform
 \cite{WfmMathWorld-HilbertTransform,Davies-IntegralTfs}, 
 is an integral transform that 
 convolves with the step function $h(t)$.
This step $h(t)$ establishes or enforces the one sided nature
 of effects that are generated causally:

 (a) 
 A cause at $t=t_0$
 can only have effects $R(t)$ that appear for times $t \ge t_0$, 
 so that multiplying by $h(t-t_0)$ has no effect, 
 i.e. $R(t) = R(t) h(t-t_0)$, as on fig. \ref{fig-RQ}.

 (b) 
 A historical record
 $S(t)$ at a time $t_0$ only contains data on the past ($t \le t_0$), 
 i.e. $S(t) = S(t) h(t_0-t)$, as on fig. \ref{fig-thepast}.

 (c) 
 A linear response function $u(t) = u(t) h(t)$, 
 where $R(t) = u(t) \convol Q(t) = \int_0^\infty u(t') Q(t-t') dt'$ 
 only depends on past values of $Q(t)$. 
 The Fourier convolution theorum 
 then tells us that 
 $\tilde{R}(\omega)=\tilde{u}(\omega) \tilde{Q}(\omega)$.

\item
The Fourier transform %\cite{Davies-IntegralTfs}
 is based on the exponential $e^{-\imath \omega t}$
 (see eqn. \eqref{eqn-FourierTf}).
It is widely used 
 to re-represent 
 a response $R(t)$, 
 time history $S(t)$, 
 or response function $u(t)$ 
 as a spectrum.
This requires that the Fourier transforms of $R$, $S$, or $u$
 are well behaved enough to exist, 
 which usually requires them to be normalizable
 and to vanish fast enough as $\omega \rightarrow \infty$.

\end{enumerate}

The Hilbert and Fourier transforms
 combine to turn time-domain ``step'' restrictions 
 on the real-valued $X(t) \in \{ R, u, S \}$ 
 into spectral constraints on the complex $\tilde{X}(\omega)$.
Following a well known theorem of Titchmarsh \cite{Titchmarsh-FI,KKinOMR},
 we can state that
 for some causal (i.e. step-like) function $X(t)$
 which depends only on the past (i.e. $t<0$), 
 the real and imaginary parts of its frequency spectrum
 are connected by
~ 
\begin{align}
  \tilde{X}(\omega)
&=
  \frac{\sigma}{\imath \pi}
  \mathscr{P}
  \int_{-\infty}^{+\infty}
    \frac{\tilde{X}(\omega')}
         {\omega'-\omega}
    d\omega'
,
 \label{eqn-tKK-standard}
\end{align}
 where the prefactor $\imath$ serves
 to cross-link the real and imaginary parts of $\tilde{X}$; 
 often the two parts are written explicitly as 
\begin{align}
  \RealPart\left[\tilde{X}(\omega)\right]
&=
  \frac{\sigma}{\pi}
  \mathscr{P}
  \int_{-\infty}^{+\infty}
    \frac{\ImagPart\left[\tilde{X}(\omega')\right]}
         {\omega'-\omega}
    d\omega'
,
 \label{eqn-tKK-standard-Re}
\\
  \ImagPart\left[\tilde{X}(\omega)\right]
&=
 -
  \frac{\sigma}{\pi}
  \mathscr{P}
  \int_{-\infty}^{+\infty}
    \frac{\RealPart\left[\tilde{X}(\omega')\right]}
         {\omega'-\omega}
    d\omega'
.
 \label{eqn-tKK-standard-Im}
\end{align}
The preferred direction for ``the past''
 is set by $\sigma$, 
 for a response $R$ or $u$, 
 we set $\sigma=+1$; 
 for a historical record $S$ we use $\sigma=-1$.
The operator $\mathscr{P}$ takes the principal part 
 \cite{WfmMathWorld-PrincipalValue} of the integral, 
  returning what we would get for the integral 
  if those points at which the integrand diverges were 
  skipped.
Both eqns. \eqref{eqn-tKK-standard-Re} and \eqref{eqn-tKK-standard-Im}
 thus inform us as to the spectral effect of temporal causality.
However,
 it is not necessary to understand the mathematics they rely on --
 i.e. integral transforms 
 and (complex) contour integration -- 
 in order to appreciate their meaning.

What the KK relations tell us is that 
 local properties are tied to global ones, 
 as noted later on fig. \ref{fig-simple-Lorentz-omega}.
If the complex valued $\tilde{X}(\omega) = $ represents a response function,
 then the real part ($\RealPart\left[\tilde{X}(\omega)\right]$)
 is the dispersion --
 in optics, 
 this might be the refractive index change $n(\omega)-1$, 
 whose frequency dependence in glass 
 gives rise to differing phase and group velocities \cite{Kinsler-2009pra}
 for different colours, 
 as well as different angles of refraction.
The imaginary part ($\ImagPart\left[\tilde{X}(\omega)\right]$)
 is then the loss or absorption, 
 or, 
 in amplifying media (as in a laser), 
 the gain.
We can then see that 
 the KK relation eqn. \eqref{eqn-tKK-standard-Re} links 
 the global loss properties ($\ImagPart[\tilde{X}(\omega')]$
 for all $\omega'$)
 to the response at a specific frequency ($\RealPart[\tilde{X}(\omega)]$).
Likewise, 
 eqn. \eqref{eqn-tKK-standard-Im} links 
 the global dispersive properties ($\RealPart[\tilde{X}(\omega')]$, 
 for all $\omega'$)
 to losses at a specific frequency ($\ImagPart[\tilde{X}(\omega)]$).
More generally, 
 any effect requiring complicated behaviour
 for the real part of the spectra 
 (usually described as ``dispersion'')
 usually has an associated imaginary component, 
 which for passive systems can usually be interpreted as loss.
This is the origin of commonly made 
 (and not always true) statements along the lines of
 ``dispersion requires (or induces) loss'' 
 \cite{Stockman-2007prl,Kinsler-M-2008prl}.

%
% =======================================================================
\section{A driven, damped oscillator}\label{S-Pendulumn}

\begin{figure}
\includegraphics[angle=-0,width=0.72\columnwidth]{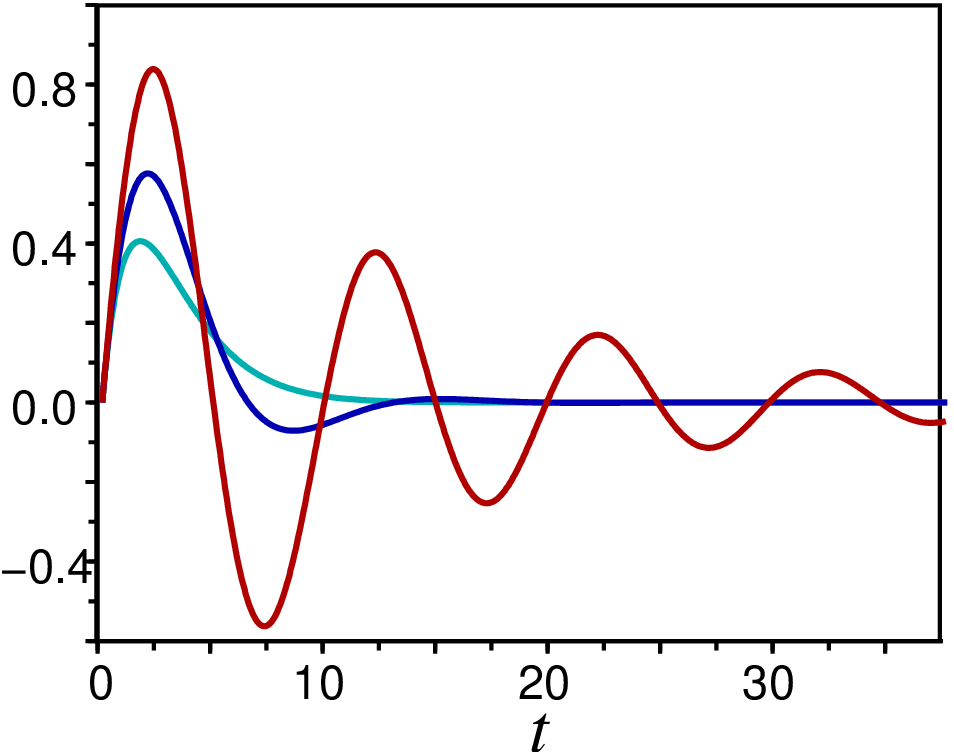}
\caption{
Typical temporal responses 
 (e.g. either $x(t)$ or ${P}(t)$) 
 to an impulsive driving force.
 for a damped oscillator
 in the underdamped (oscilliatory) regime.
The initial ramp-like response can be seen 
 close to the vertical axis near $t=0$.
}
\label{fig-simple-Lorentz-t}
\end{figure}

Let us now consider a causal response
 more complicated than the simple case shown in eqn. \eqref{eqn-basicRQ}, 
 but both simpler and more specific
 than the very general form of eqn. \eqref{eqn-multiRQ}.
An ideal example is that of a driven, damped oscillator
 such as a mass on a spring 
 \cite{WfmMathWorld-DampedSHM,Wikipedia-DampedSHM}, 
 whose temporal differential equation matches that 
 often used in electromagnetism 
 to model the Lorentz response
 in a dielectric medium \cite{RMC,Jackson-ClassicalED}.
For the mass on a spring, 
 with a spring constant $k$, 
 we have a resonant frequency of $\omega_0 = \sqrt{k/m}$, 
 and a friction (or ``loss'') parameter $\gamma$; 
 likewise the Lorentz response also has a resonant frequency and loss.
Given these parameters, 
 the displacement of the pendulumn bob $x(t)$
 (or dielectric polarization $\Vec{P}$) 
 could then be affected
 by the driving force per unit mass $F(t)/m$
 (or electric field $\Vec{E}$)
 according to equations of the form
~
\begin{align}
  \partial_t^2 x(t) 
 +
  \gamma \partial_t x(t)
 +
  \omega_0^2 x(t)
&=
  F(t) / m
,
\label{eqn-std-Pendulumn}
\\
  \partial_t^2 \Vec{P}(t) 
 +
  \gamma \partial_t \Vec{P}(t)
 +
  \omega_0^2 \Vec{P}(t)
&=
  \alpha \epsilon_0 \Vec{E}(t)
.
\label{eqn-std-Lorentz0}
\end{align} 
Here a delta function impulse in force $F(t) = p_0 \delta(t)$
 does not induce an initial step change in position, 
 but in velocity $\partial_t x(t) \simeq p_0/m$; 
 with the likewise \emph{initial} response of a linear 
 (or ramp-like)
 change in position, 
 with $x(t) \simeq t p_0/m$.
In the same way, 
 in an electromagnetic Lorentz dielectric medium,
 an impulsive $\Vec{E}(t) = \Vec{j}_0 \delta(t)$ 
 gives rise to an initial step change
 in polarization current 
 $\partial_t \Vec{P}(t) \simeq \alpha \epsilon_0 \Vec{j}_0 $, 
 and a concomittant ramp/linear change in polarization initially,
 i.e. $\Vec{P}(t) \simeq \alpha \epsilon_0 t \Vec{j}_0$.
Fig. \ref{fig-simple-Lorentz-t} shows some typical oscillating
 (under-damped) time responses to an impulsive driving force.

\begin{figure}
\includegraphics[angle=-0,width=0.72\columnwidth]{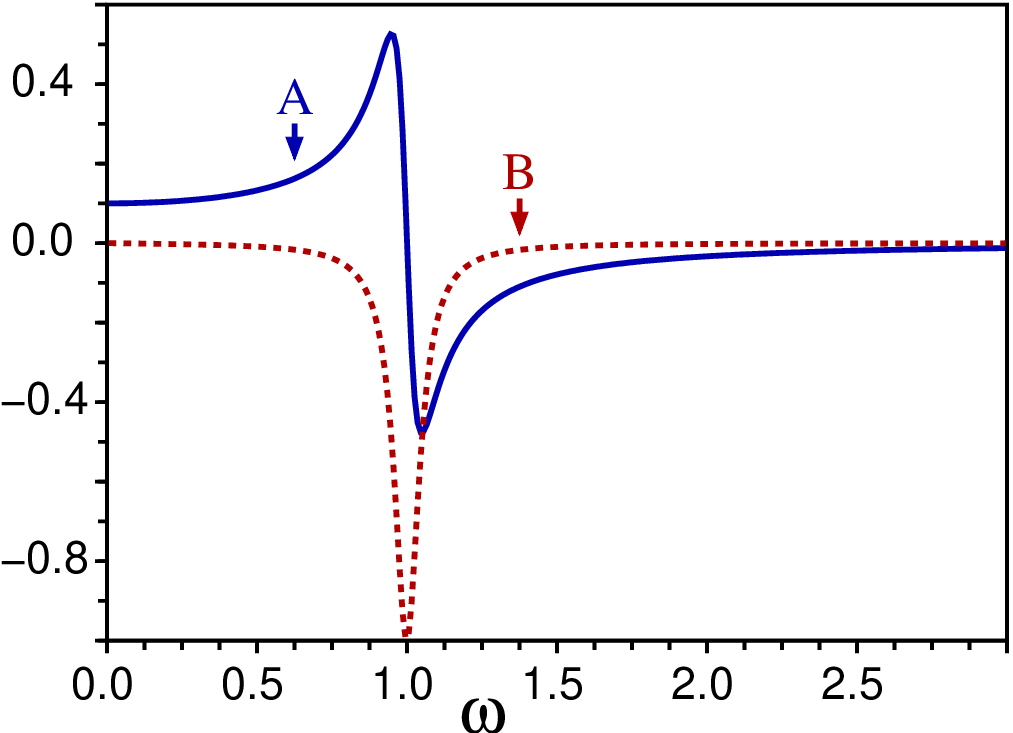}
\caption{
A typical spectral response $\tilde{u}(\omega)$ for the 
 damped ocillator model.
The solid line shows the real part of the response, 
 the dashed line the imaginary part.
The real part of the response at any point 
 (e.g. A at $\omega \simeq 0.6$)
 depends on an integral of the imaginary part over all frequencies --
 i.e. over the whole dashed line, $\omega' \in [0, \infty ]$.
Similarly, 
 likewise the imaginary part of the response at any point
 (e.g. B at $\omega \simeq 1.4$)
 depends on an integral of the real part over all frequencies --
 i.e. over the whole solid line, $\omega' \in [0, \infty ]$.
}
\label{fig-simple-Lorentz-omega}
\end{figure}

Both eqns. \eqref{eqn-std-Pendulumn} and \eqref{eqn-std-Lorentz0} 
 are linear, 
 so that the model can also be expressed in terms of
 a response function $u(t)$ --
 e.g. for the dielectric, 
 we would have that 
 $\Vec{P}(t) = \int_0^\infty u(t') \Vec{E}(t-t') dt'$.
We can then Fourier transform this, 
 and when the transform of $u(t)$ is denoted $\tilde{u}(\omega)$,
 we have that 
 $\tilde{\Vec{P}}(\omega) = \tilde{u}(\omega) \tilde{\Vec{E}}(\omega)$.
Since the Fourier transform of eqn. \eqref{eqn-std-Lorentz0} is
~
\begin{align}
  \left[
   -
    \omega^2
   -
    \imath \gamma \omega
   +
    \omega_0^2
  \right]
  \tilde{\Vec{P}}(\omega) 
&=
  \alpha \epsilon_0 \tilde{\Vec{E}}(\omega) 
,
\label{eqn-std-Lorentz-transformed}
\end{align} 
 the spectral response $\tilde{u}(\omega)$ is then easily obtained,
 being
~
\begin{align}
  \tilde{u}(\omega)
&=
  \frac{ - \alpha \epsilon_0 }
  {
    \omega^2 - \omega_0^2 + \imath \gamma \omega
  }
,
\label{eqn-std-Lorentz-w}
\end{align}
 whose real and imaginary parts
 are shown on fig. \ref{fig-simple-Lorentz-omega}.
In an electromagnetic dielectric $\tilde{u}(\omega)$
 is related to the refractive index $n$
 by $n^2 = 1 + \tilde{u}(\omega)/\epsilon_0$ 
 \cite{Kinsler-2009pra}.
We can see from eqn. \eqref{eqn-std-Lorentz-w}
 that the real part of $\tilde{u}(\omega)$
 has a frequency dependent variation 
 with an explicit dependence on the loss parameter $\gamma$.
Likewise, 
 the loss-like part of the response, 
 i.e. the imaginary part of $\tilde{u}(\omega)$, 
 has an explicit dependence on frequency.

Not all models that we might derive or write down will be causal.
A contemporary example is the F-model \cite{Pendry-HRS-1999ieeemtt}, 
 which is used to describe the magnetic response
 (i.e. the magnetization $\Vec{M}$)
 of the split-ring resonators (SRRs)
 often used in electromagnetic metamaterials.
While similar to the Lorentz model, 
 its driving term is instead the second derivative
 of the magnetic $\Vec{H}$ field.
At frequencies relevant to typical applications, 
 the F-model works well; 
 but for high frequencies where $\omega \gg \omega_0$,
 the associated short wavelengths cause
 the approximations used to derive it to fail.
The F-model follows
~
\begin{align}
  \partial_t^2 \Vec{M}(t) 
 +
  \gamma \partial_t \Vec{M}(t)
 +
  \omega_0^2 \Vec{M}(t)
&=
  \alpha \partial_t^2 \Vec{H}(t)
.
\label{eqn-std-Fmodel}
\end{align} 
The spectral response based on $\Vec{M}(t) = u_F(t) \convol \Vec{H}(t)$
 or $\tilde{\Vec{M}}(\omega) = u_F(\omega) \tilde{\Vec{H}}(\omega)$ is
~
\begin{align}
  \tilde{u}_F(\omega)
&=
    \frac{\alpha \omega^2}
         {\omega_0^2 - \omega^2 - \imath \omega \gamma_\mu}
.
\label{eqn-mu-nonLorentzian}
\end{align}
This isn't KK causal,
 because eqn. (\ref{eqn-std-Fmodel})
 violates the criterion of having lower-order time derivatives
 on the right as compared to the left, 
 and (therefore) eqn. (\ref{eqn-mu-nonLorentzian})
 fails the test imposed by the Kramers Kronig relations:
 the response remains finite at arbitrarily high frequencies.
{More descriptively, 
 we can say that
 ever higher frequency components of $\Vec{H}$ and $\Vec{M}$ 
 become ever more synchronized; 
 and synchronized quantities are related 
 not by cause-and-effect, 
 but with an equality.}

We can, 
 however, 
 fix the F-model 
 by defining a new auxiliary field $\Vec{K} = \Vec{M} - \alpha \Vec{H}$.
{This separates the changes in $\Vec{M}$ \emph{caused} by $\Vec{H}$,
 from those \emph{synchronized} with $\Vec{H}$,
 by allowing us to merge the two highest-order derivative terms.
The resulting explicitly causal F-model, 
 where $\Vec{H}$ and $\partial_t \Vec{H}$ cause changes in $\Vec{K}$,}
 is now compatible with eqn. (\ref{eqn-multiRQ}),
 and follows %the differential equation
~
\begin{eqnarray}
  \partial_t^2 \Vec{K}(t) 
 +
  \gamma \partial_t \Vec{K}(t)
 +
  \omega_0^2 \Vec{K}(t)
&=
 -
  \alpha \gamma \partial_t \Vec{H}(t)
 -
  \alpha \omega_0^2 \Vec{H}(t)
,
\label{eqn-std-Fmodel-variant}
\end{eqnarray}
which has a spectral response $\tilde{u}_K$ based on 
 $\Vec{K}(t) = u_K(t) \convol \Vec{H}(t)$
 or $\tilde{\Vec{K}}(\omega) 
        = \tilde{u}_K(\omega) \tilde{\Vec{H}}(\omega)$.
It vanishes correctly at high frequencies, 
 and is
~
\begin{eqnarray}
  \tilde{u}_K(\omega)
&=
 -
  \alpha
    \frac{- \imath \gamma \omega + \omega_0^2}
         {\omega_0^2 - \omega^2 - \imath \omega \gamma_\mu}
.
\label{eqn-mu-nonLorentzian-variant}
\end{eqnarray}
We can use either of eqn. (\ref{eqn-std-Fmodel-variant})
 or (\ref{eqn-mu-nonLorentzian-variant})
 to evaluate $\Vec{K}$, 
 after which we can extract $\Vec{M}$ and find $\Vec{B}$ using
$ \Vec{M} = \Vec{K} + \alpha  \Vec{H}$
 and the constitutive relation
$ \mu_0^{-1} \Vec{B} = \Vec{H} + \Vec{M} 
= \Vec{K} + \left(1+\alpha\right)  \Vec{H}
$.
Although re-expressing the F-model in terms of $\Vec{K}$
 does not alter the many physical approximations 
 made in its derivation,
 by shifting the non-causal part into
 the constitutive relation for $\Vec{B}$ (an equality),
 it ensures that the dynamic response in 
 eqn. (\ref{eqn-std-Fmodel-variant})
 is now explicitly causal.

%
% =======================================================================
\section{Maxwell's equations}\label{S-Maxwell}

The curl Maxwell's equations
 control the behaviour
 of the electric and electric displacement fields 
  $\Vec{E}(\Vec{r},t)$ and $\Vec{D}(\Vec{r},t)$, 
 and the magnetic and magnetic induction fields 
  $\Vec{B}(\Vec{r},t)$ and $\Vec{H}(\Vec{r},t)$; 
 and depend on a current density $\Vec{J}(\Vec{r},t)$.
Although usually written with the curl operator
 on the LHS \cite{RMC,Jackson-ClassicalED}, 
 our simple causal model of eqn. \eqref{eqn-basicRQ}
 leads instead to
~
\begin{align}
  \partial_t \Vec{D}
&=
  \grad \cross \Vec{H} - \Vec{J}
, 
\quad
&
  \partial_t \Vec{B}
&=
 -
  \grad \cross \Vec{E}
.
\label{eqn-basiccurlMaxwell}
\end{align}
These otherwise independent pairs $\Vec{E}, \Vec{B}$ 
 and $\Vec{D}, \Vec{H}$ \cite{HehlObukhov}
 are connected together by the constitutive relations 
 involving the dielectric polarization $\Vec{P}$
 and magnetization $\Vec{M}$ of the background medium, 
 which are
~
\begin{align}
  \Vec{D}=\epsilon_0 \Vec{E} + \Vec{P}
,
 &&
  \Vec{B} = \mu_0 \Vec{H} + \mu_0 \Vec{M}
,
\label{eqn-basicConstitutive}
\end{align}
 and are subject to the constraint imposed by 
 the divergence Maxwell's equations, 
 which depend on the free electric charge density $\rho(\Vec{r},t)$ and 
 the zero magnetic charge density, 
 and are
~
\begin{align}
  \grad \cdot \Vec{D} = \rho %/\epsilon_0,
 &&
  \grad \cdot \Vec{B} = 0
.
\end{align}

Perhaps surprisingly, 
 the causal nature of Maxwell's equations remains a subject of
 debate (see e.g. \cite{Savage-2011arXiv-cem,Jefimenko-2004ejp}).
Nevertheless,
 the curl Maxwell's equations must be causal in the ``step'' KK sense:
 eqns. (\ref{eqn-basiccurlMaxwell}) have the same form
 as our simple causal model in eqn. \eqref{eqn-basicRQ}.
The curl operators,
 by specifying how the spatial profile of $\Vec{H}$ and $\Vec{E}$
 drive changes in $\Vec{D}$ and $\Vec{B}$, 
 turns them into wave equations when combined
 with the constitutive relations.
Take as a starting point the case in vacuum, 
 where $\Vec{P}$ and $\Vec{M}$ are both zero, 
 so that $\Vec{D}=\epsilon_0 \Vec{E}$ and $\Vec{B} = \mu_0 \Vec{H}$.
Then 
 we can then rewrite eqns. \eqref{eqn-basiccurlMaxwell}
 solely in terms of any pair of electric-like ($\Vec{E}$ or $\Vec{D}$)
 and magnetic-like ($\Vec{H}$ or $\Vec{B}$) fields; 
 e.g.
~
\begin{align}
  \epsilon_0 \partial_t \Vec{E}
&=
  \grad \cross \Vec{H} - \Vec{J}
, 
\quad
&  \mu_0 \partial_t \Vec{H}
&=
 -
  \grad \cross \Vec{E}
,
\label{eqn-vaccumcurlMaxwell}
\end{align} 
 which can be combined into second order forms for $\Vec{E}$ or $\Vec{B}$, 
 e.g. $\epsilon_0 \mu_0 \partial_t^2 \Vec{E} = 
  - \grad \times \grad \times \Vec{E} - \mu_0 \partial_t \Vec{J}$.
While these vacuum Maxwell's equations are self-evidently KK causal,
 more generally,
 the background medium for the electromagnetic fields 
 can have non-trivial and dynamical responses
 to those fields encoded in $\Vec{P}$ and $\Vec{M}$.
To avoid specifying particular response models,
 I represent all possible causal differential equations
 for $\Vec{P}$ and $\Vec{M}$
 (compatible with eqn. \eqref{eqn-multiRQ})
 with the notation
~
\begin{align}
  \delta_t \Vec{P} = \Vec{f}(\cdot),
&&
  \delta_t \Vec{M} = \Vec{f}(\cdot)
.
\end{align}
As an example, 
 $\delta_t \Vec{P} = \Vec{f}(\Vec{E})$ might be used to represent 
 eqn. \eqref{eqn-std-Lorentz0}, 
 the Lorentz response in a dielectric.

A straightforward expression of 
 Maxwell's equations
 that emphasizes their causal nature 
 is achieved by insisting that 
 any given field should not appear on both
 the LHS and RHS of the equations.
Thus every field that has a dynamical response
 (i.e. is modelled by a temporal differential equation)
 can be updated simultaneously.
There is no need to follow some specified sequence, 
 although that can be useful, 
 as in e.g. finite element simulations \cite{Taflove-FDTD}.
We can even do this even whilst 
 incorporating magneto-electric material responses, 
 where the electric field affects the magnetization, 
 or the magnetic field affects the dielectric polarization.
Maxwell's equations, 
 written to fit these criteria, 
 are
~
\begin{align}
  \partial_t \Vec{D} &= +\grad \cross \Vec{H} - \Vec{J}, \quad
&
  \delta_t \Vec{M} &= \Vec{f}(\Vec{H},\Vec{E})
\label{eqn-maxwell-cleancausal-DM}
\\
\textrm{and} \qquad
  \partial_t \Vec{B} &= -\grad \cross \Vec{E}, \quad
 & 
  \delta_t \Vec{P} &= \Vec{f}(\Vec{E},\Vec{H})
\label{eqn-maxwell-cleancausal-BP}
.
\end{align}
Although interdependent, 
 these equations remain explicitly KK causal in the sense that 
 $\Vec{H}$ and $\Vec{E}$ are uniquely defined as causes, 
 and $\Vec{D}, \Vec{P}$ and $\Vec{B}, \Vec{M}$ are affected by those causes 
 (i.e. show ``effects'').
Further,
 we cannot regard the displacement current $\partial_t \Vec{D}$
 (or indeed its magnetic counterpart $\partial_t \Vec{B}$)
 as ``causes'' in the manner reviewed by Heras \cite{Heras-2011ajp}; 
 these changes in $ \Vec{D}$ and $\Vec{B}$ are instead \emph{effects}.

In typical non magneto-electric cases,
 where $\delta_t \Vec{M} = \Vec{f}(\Vec{H})$
 and $\delta_t \Vec{P} = \Vec{f}(\Vec{E})$,
 the two equation sets \eqref{eqn-maxwell-cleancausal-DM}
 and \eqref{eqn-maxwell-cleancausal-BP} are independent of one another.
This further possible separation is the reason for associating 
 the equation for $\Vec{M}$ with that for $\Vec{D}$ 
 in eqn. \eqref{eqn-maxwell-cleancausal-DM}, 
 and associating that for $\Vec{P}$ with $\Vec{B}$ 
 in eqn. \eqref{eqn-maxwell-cleancausal-BP}.

Once the RHS's of eqns. \eqref{eqn-maxwell-cleancausal-DM}
 and \eqref{eqn-maxwell-cleancausal-BP} have been evaluated, 
 the LHS's can be integrated directly in an explicitly KK causal manner -- 
 the ``cause'' fields $\Vec{E}, \Vec{H}$
 have had effects on $\Vec{D}, \Vec{P}$ and $\Vec{B}, \Vec{M}$.
Then, 
 the usual constitutive relations can be rearranged 
 to connect the fields according to
~
\begin{align}
  \Vec{E} = \epsilon_0^{-1} \left[ \Vec{D} - \Vec{P} \right]
,
 &&
  \Vec{H} = \mu_0^{-1} \Vec{B} - \Vec{M}
,
\end{align}
 to 
 allow us to directly update the  the ``cause'' fields $\Vec{E}$ and $\Vec{H}$.

Note that if the evolution of $\Vec{M}$ or $\Vec{P}$
 were (e.g.) to be written as dependent on $\Vec{D}$ or $\Vec{B}$, 
 (so that $\delta_t \Vec{M} = \Vec{f}(\Vec{B},\Vec{D})$), 
 then there is no longer a perfect separation between ``cause fields'' 
 and ``effect fields''.
Nevertheless, 
 such a rewriting will not violate causality, 
 since the differential equations still have the correct form.

We can easily replace the abstract current density $\Vec{J}$
 by incorporating the motion of a particles of mass $m_j$
 and charge $q_j$ at position $\Vec{x}_j(t)$ with velocity $\Vec{v}_j(t)$, 
 by using additional causal equations 
~
\begin{align}
  \partial_t \Vec{v}_j(t)
&=
  \frac{q_j}{m_j} 
  \left[
    \Vec{E}(\Vec{x}_j(t),t)
   +
    \Vec{v}_j(t) \cross \Vec{B}(\Vec{x}_j(t),t)  
  \right]
\\
  \partial_t \Vec{x}_j(t)
&=
  \Vec{v}_j(t)
,
\end{align}
 along with the connection between the electric current density
 and the particle motion, 
 and the charge density,
 which are
~
\begin{align}
  \Vec{J}(\Vec{r},t) 
&=
  \sum_j
  q_j \delta(\Vec{r} - \Vec{x}_j(t)) \Vec{v}_j (t)
\\
  \rho(\Vec{r},t)  
&=
  \sum_j q_j \delta(\Vec{r} - \Vec{x}_j(t))
.
\end{align}

Of course, 
 how light changes its propagation properties inside some material 
 is not always described using charges, 
 currents, 
 or polarization and/or magnetization.
Instead we often use its refractive index $n(\omega)$,
 which has two distinct contributions.
First, 
 there is the vacuum or spacetime metric component,
 which is unity at all frequencies; 
 it is this part that was historically interpreted as 
 an instantaneously responding electromagnetic ``ether''.
Second, 
 there is that due to the polarization and magnetization responses
 (i.e. either $n(\omega)-1$ or $n^2(\omega)-1)$ \cite{Kinsler-2009pra}), 
 and it is this second component which 
 causality (and the KK relations)
 insist should vanish at high frequencies.

Finally, 
 note that this approach can be also be applied to Faraday's Law, 
 which result in some perhaps surprising conclusions
 \cite{Kinsler-2017arxiv-faradin}.

%
% =======================================================================
\section{Waves and signals}\label{S-waves}

Causality can be broadened 
 to include the notion of ``signal causality'' --
 where signals must pass between 
 two separate points for influences to be felt.
To analyse this problem, 
 I take the postion that signals are sent using waves, 
 and that waves follow wave equations.
This is motivated by the important case of lightspeed signals, 
 as propagated according to Maxwell's equations.
We will see that light propagation is not only ``time-step'' KK causal
 as discussed in section \ref{S-Maxwell}, 
 but ``wave-cone'' (light-cone) causal as well.

To start,
 let us convert our simple causal differential equation \eqref{eqn-basicRQ}
 into a wave equation by augmenting the cause $Q(x,t)$
 by a spatial derivative of the effect $R$
 and a wave speed $c$,
 e.g.
~
\begin{align}
  \partial_t R(x,t) %{\pm}
&=
 \pm
  c
  \partial_x R(x,t)
 +
  Q(x,t)
,
\label{eqn-wave-basic}
\end{align} 
where $\partial_x$ is just the spatial derivative $d/dx$.
Thus we can see that waves are (at least in part) their own causes, 
 where in particular the ``cause'' of changes in profile 
 is their own pre-existing spatial modulation,
 in addition to any source-like causes $Q$.
To obtain bi-directional propagation we need either higher order derivatives
 or multi-component waves, 
 e.g. 
 in three dimensions we need at least one scalar
 and one vector component.
As for linearized acoustic waves
 in pressure and fluid velocity \cite{MorseIngard-Acoustics}, 
 % kappa = compressibility, rho=density; see sec 6.2 p241-243
 we could link a scalar $R$ and vector $\Vec{P}$
 with a pair of differential equations:
~
\begin{align}
  \partial_t R(\Vec{r},t)
&=
  \kappa^{-1}
  \grad 
  \cdot
  \Vec{P}(\Vec{r},t)
 +
  Q_R(\Vec{r},t)
,
\label{eqn-wave-basic-R}
\\
  \partial_t \Vec{P}(\Vec{r},t)
&=
  \rho^{-1}
  \grad 
  R(\Vec{r},t)
 +
  \Vec{Q}_P(\Vec{r},t)
.
\label{eqn-wave-basic-P}
\end{align} 
These can be combined into either of two second order forms,
 the scalar one being the simplest, 
 which with $Q(\Vec{r},t) 
          = \partial_t Q_R(\Vec{r},t)
           + \kappa^{-1} \grad \cdot \Vec{Q}_P(\Vec{r},t)$
 and $c^2 =  1/\rho\kappa$
 is 
~
\begin{align}
  \partial_t^2 R(\Vec{r},t) 
&=
  c^2  \grad^2 R(\Vec{r},t)
 +
  Q(\Vec{r},t)
\label{eqn-wave-basic2nd-R}
.
\end{align} 
This is explicitly KK causal, 
 since the LHS has a second order time derivative, 
 whilst the RHS has at most only a single time derivative inside $Q$.
Many other wave models are possible, 
 notably the transverse electromagnetic waves
 generated by Maxwell's equations (see section \ref{S-Maxwell}), 
 for which a very general second order form
 comparable to eqn. \eqref{eqn-wave-basic2nd-R} exists
  \cite{Kinsler-2010pra-fchhg}; 
 as well as a first order bidirectional version
  \cite{Kinsler-2010pra-dblnlGpm}.
The hyperbolic form of these second order wave equations
 gives rise to ``wave-cones'' analogous to the light cones 
 discussed in relativistic scenarios.

Most simple linear waves
 follow second order equations like eqn. \eqref{eqn-wave-basic2nd-R}, 
 although different physical properties
 are denoted by $R$ and $Q$ in the different cases.
Such waves provide a useful basis for analysing
 causal signalling, 
 since they can be used to establish KK relations
 that constrain spectra in both space and time.
It is important to choose as a basis those waves travelling
 at the fastest relevant signal speed.
Of course the speed of light can always fulfill this role, 
 but there may be cases when a lower speed limit is appropriate
 \cite{Todd-M-2017fp}.

\begin{figure}
\includegraphics[angle=-0,width=0.60\columnwidth]{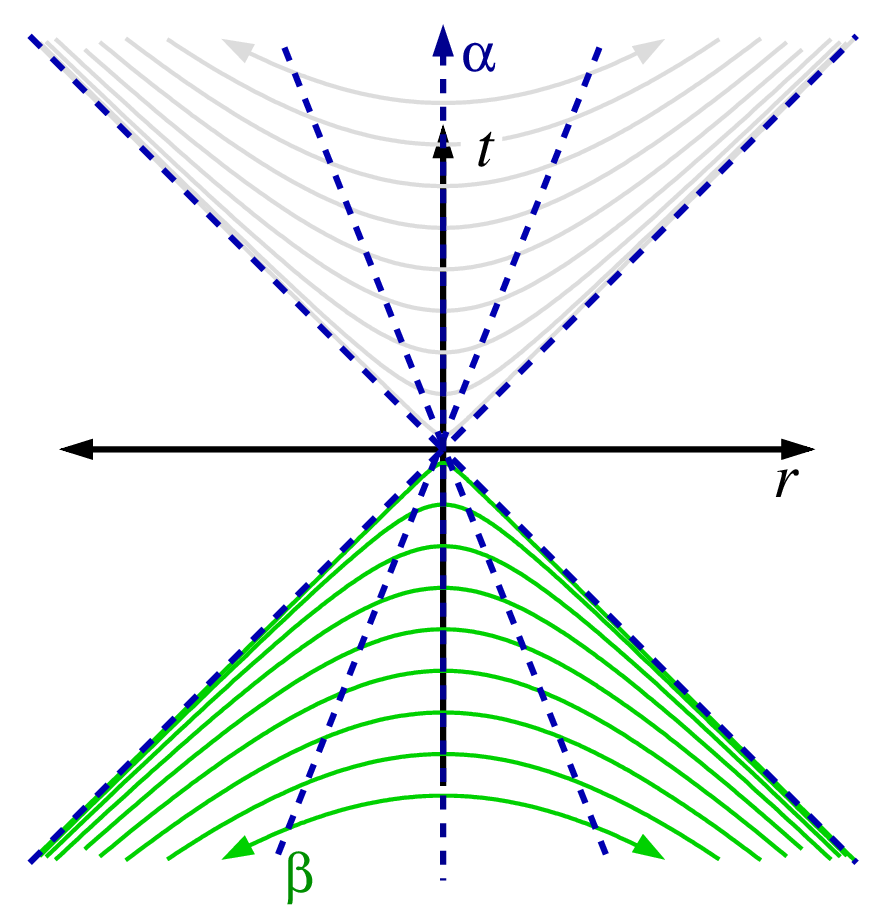}
\caption{
The wave-cone coordinates $\alpha$ and $\beta$
 plotted on the usual $t$ and $r$ axes.
For fixed $\alpha$, 
 $\beta$ varies along hyperbol{\ae}, 
 for fixed $\beta$, 
 $\alpha$ varies along (dashed) straight lines in the $r,t$ plane.
}
\label{fig-lightcone-hyp}
\end{figure}

In special relativity, 
 one important frame independent quantity 
 for any two spacetime points 
 (i.e. here the origin and $\Vec{r},t$)
 is the interval $\alpha$ between them \cite{Schutz-FCRelativity}.
It also plays a crucial role in spacetime causality,
 since it is equally useful for waves in an isotropic material medium, 
 as long as we remain in the rest frame of that medium. 
The interval is
~
\begin{align}
  \alpha 
=
  \sqrt{c^2 t^2 - \Vec{r} \cdot \Vec{r}} 
=
  +\sqrt{\alpha^2} . \sgn(t)
.
\label{eqn-wave-alpha}
\end{align}
Here I have used the sign of $t$ to distinguish positive intervals
 (between now and points in the future wave-cone) 
 and negative intervals (between now and points in the past wave-cone); 
 imaginary values of $\alpha$, 
 resulting from $c^2 t^2 < \Vec{r} \cdot \Vec{r}$, 
 are excluded.
We also need a rapidity coordinate
~
\begin{align}
  \beta = \arctanh(|\Vec{r}|/ct)
,
\label{eqn-wave-beta}
\end{align}
which indicates the average speed $v=|\Vec{r}|/t$ 
 required to travel between a point $\Vec{r},t$ and the origin.
How these relate to $r$, $t$ coordinates
 is shown on fig. \ref{fig-lightcone-hyp}.
The inverse transformation is
~
\begin{align}
  ct
&=
  \alpha ~ \cosh(\beta)
,
\\
  \left| \Vec{r} \right|
&=
  \alpha ~ \sinh(\beta)
.
\end{align}
In special relativity, 
 where $c$ is the speed of light, 
 $\alpha$ and $\beta$ are related to the Rindler coordinates, 
 but are for timelike intervals rather than spacelike ones.
To complete the coordinate set, 
 we need the two polar angles that enable us to reconstruct
 the individual position coordinates $\Vec{r} = (r_x,r_y,r_z)$
 from their magnitude $r =  | \Vec{r}|$; 
 using $\theta = \tan^{-1} ( r_y / r_x)$
 and
 $\phi   = \cos^{-1} (r_z / \left| \Vec{r} \right|)$. 
For notational convenience, 
 we can combine $\beta$ with $\theta, \phi$
 to form a single rapidity vector 
 $\Vec{\beta}= (\beta, \theta, \phi)$; 
 we might also define pseudo-Cartesian coordinates with
 $\beta_x = \beta \cos \theta \cos \phi$, 
 $\beta_y = \beta \cos \theta \sin \phi$, 
 and
 $\beta_z = \beta \sin \theta$
 to give $\Vec{\beta}= (\beta_x, \beta_y, \beta_z)$.

In the 1D linear case with $r \equiv x$ and $S(r,t)  = R(x,t)$, 
 or the spherically symmetric case with ${S}(r,t)  = r R(r,t)$, 
 the wave equations in both sets of coordinates are
~
\begin{align}
  \partial_t^2 {S}(r,t) 
&= 
  c^2
  \partial_r^2 {S}(r,t)
 + 
  {Q}'(r,t)
,
\label{eqn-wave-radial-tr}
\\
  \alpha^2 \partial_\alpha^2 {S}(\alpha,\beta) 
&= 
  \partial_\beta^2 {S}(\alpha,\beta)
 + 
  {Q}'(\alpha,\beta)
,
\label{eqn-wave-radial-ab}
\end{align}
 where the source terms are written simply as a ${Q}'$, 
 even though they vary between cases.
The critical difference between the $r,t$ coordinates 
 and the $\alpha, \Vec{\beta}$ ones is that for the latter choice, 
 the edge of the wave-cone is coincident with $\alpha=0$
 and is never crossed. 
Thus ``spacelike'' intervals are not covered
 by these hyperbolic coordinates.
As a result,
 only the wave interval $\alpha$ has causal restrictions.
And whilst eqn. \eqref{eqn-wave-radial-tr}
 is explicitly time-step or KK causal, 
 eqn. \eqref{eqn-wave-radial-ab} is something more --
 it is wave-cone causal
 (or, as named below, $\Omega K$KK causal).

When considering the wave equations \eqref{eqn-wave-radial-tr}
 and \eqref{eqn-wave-radial-ab},
 note that the eigenfunctions
 of $\partial_t^2$, $\partial_r^2$, or $\partial_\beta^2$ 
 are just plane waves (e.g.) $\exp(\pm \imath K \beta)$
 with eigenvalues $-K^2$. %$(\pm \imath K)^2$.
In contrast 
 those of $\alpha^2 \partial_\alpha^2$ are powers of $\alpha$, 
 where eigenfunctions $\alpha^{1/2} \alpha^{\pm \imath \Pi}$ %\Omega}$
 have eigenvalues $-(\Pi^2+1/4)$.
This means that while the Fourier transform is the natural choice
 for a spectral transformation on $t$, $r$, or $\beta$, 
 it is not so for $\alpha$.
In fact the power-law dependence of the eigenfunctions for $\alpha$
 indicates that the Mellin transform 
 \cite{WfmMathWorld-MellinTransform,Davies-IntegralTfs}
 should be the natural choice.

Nevertheless,
 the mathematical construction of the KK relations
 depends only on having a step function along some coordinate.
Thus we can still use the Fourier transform 
 to construct spacetime KK relations, 
 and so avoid for now using the lesser known Mellin transform.
Taking the Fourier spectral counter part of $\alpha$
 to be $\Omega$, 
 we can rewrite eqn. \eqref{eqn-tKK-standard}
 and get a KK relation consistent with the signal causality 
 resulting from the wave equation
 \eqref{eqn-wave-basic2nd-R}. % and \eqref{eqn-wave-basic-P}.
Thus for a wave, 
 or some wave-causal quantity $S(\Vec{r},t)$,
 we have the $\Omega$KK relations
~
\begin{align}
  %\partial^3_\Omega
  \tilde{S}(\Omega,\Vec{\beta})
&=
  \frac{\sigma}{\imath \pi}
  \int_{-\infty}^{+\infty}
    \frac{ %\partial_\Omega^3 
           \tilde{S}(\Omega',\Vec{\beta})
         }
         {\Omega - \Omega'}
    d\Omega'
.
 \label{eqn-tzKK-u}
\end{align}

Further, 
 we can also consider the ${\beta}$-spectral version
 of eqn. \eqref{eqn-tzKK-u}; 
 where we denote the wavevector-like counterpart of the rapidity $\beta$
 by the symbol $K$.
Since ${\beta}$ is unbounded,  
 Fourier transforming $\tilde{S}(\Omega,\Vec{\beta})$ 
 into the $K$ domain gives the doubly transformed 
 $\bar{\tilde{S}}(\Omega,\Vec{K})$, 
 satisfying $\Omega K$KK relations
~
\begin{align}
  %\partial^3_\Omega
  \bar{\tilde{S}}(\Omega,\Vec{K})
&=
  \frac{\sigma}{\imath \pi}
  \int_{-\infty}^{+\infty}
    \frac{ %\partial_\Omega^3 
           \bar{\tilde{S}}(\Omega',\Vec{K})
         }
         {\Omega - \Omega'}
    d\Omega'
,
 \label{eqn-tzKK-uK}
\end{align}
 where $\Vec{K} = (K,\theta,\phi)$. 
Of course, 
 the angular coordinates $\theta,\phi$ could also be transformed
 into angular spectra, 
 or we could start with the 
 pseudo-Cartesian $\Vec{\beta} =(\beta_x, \beta_y, \beta_z)$ 
 and transform into its spectral counterpart $\Vec{K} = (K_x,K_y,K_z)$.

Looking back at the definition of $\alpha$, $\beta$
 we see that for pointlike objects, 
 the standard KK relations might still be applied, 
 since whatever variation in $\beta$ there is 
 merely compresses $t$ differently with respect to $\alpha$, 
 leaving the step function intact.
Nevertheless, 
 discrepancies between true spacetime $\Omega$KK causality
 and simple temporal KK causality will exist; 
 e.g. most simply, 
 a pointlike object moving with fixed rapidity $\beta$
 will have its frequencies scaled by a factor of $\cosh \beta$
 due to time dilation.

More remarkable is what eqn. (\ref{eqn-tzKK-uK})
 says about the spatial (wavevector) spectra. 
Only for a narrow range of fixed times in the past,
 where an object is well localized inside the wave-cone 
 (so that $\Delta r \ll r$ and $\Delta t \ll t$),
 can an ordinary spatial spectra $\bar{S}(k)$ based on an $S(r)$
 be made to match up with the (wave-cone) rapidity spectra $\bar{S}(K)$
 based on $S(\beta)$.
Nevertheless, 
 in typical, 
 non-relativistic,
 laboratory situations with lightspeed communication,
 we can quite reasonably restrict our analysis
 to that small region of agreement.
Otherwise, 
 our lack of knowledge
 about the situation outside the wave-cone, 
 and the mismatch between spatial coordinate ${r}$
 and the spatiotemporal ${\beta}$ 
 means that since $S({r})$ looks nothing like $S({\beta})$, 
 their respective spectra $\bar{S}({k})$, $\bar{S}({K})$
 are also utterly different:
 a structure periodic in $r$ 
 is not periodic in $\beta$.

%
%
% =======================================================================
\section{Simultaneity, locality, and smoothness}\label{S-simul}

Two points 
 raised by the stance of Jefimenko \cite{Jefimenko-2004ejp}
 question (a) whether  the use of local causality is appropriate, 
 and (b) if causes and effects can be simultaneous.

\noindent\emph{Local vs retarded causes:~}
The \emph{local} causality used here
 works knowing only the current state
 and current influences, 
 and makes only minimal assumptions.
In contrast, 
 some (e.g. Jefimenko \cite{Jefimenko-2004ejp}, 
  also see Heras \cite{Heras-2011ajp}) 
 prefer to  relate effects back to their original causes.
Thus the electromagnetic fields 
 would be directly obtained from the integral equations over the
 past behaviour of charges and currents
 (see eqns. (7,8) in \cite{Jefimenko-2004ejp}).
From a practical perspective, 
 this can raise difficulties:
 we often want to solve electromagnetic problems for free fields
 on the basis of some stated initial boundary conditions --
 where we do not know, 
 nor want to calculate,
 whatever sources may have been required to generate them.
KK causality, 
 being local, 
 neither knows or cares about this ``deep'' past; 
 but casuality is still enforced and remains testable.
However, 
 this deep past is not irrelevant, 
 and the assumption that fields can be related back to sources 
 is an important one \cite{Weinstein-2011mpla}.

\noindent\emph{Simultaneity \& smoothness:~}
We might take the position that having
 any part (however infinitesimal) 
 of the effect simultaneous with the cause is unsatisfactory; 
 e.g. as does Jefimenko \cite{Jefimenko-2004ejp}
 with regard to Maxwell's equations,
 and by implication even the simple eqn. \eqref{eqn-basicRQ}, 
 even though both are KK causal
 (even if they might not be ``Jefimenko causal'' as well).
Mathematically, 
 the step response can be made non-simultaneous
 by use of $t>t_0$ rather than $t\ge t_0$, 
 but since this has negligible physical consequences, 
 we might prefer to remove the simultaneity
 by demanding a smooth response.
This can be instituted by replacing the LHS of eqn. \eqref{eqn-basicRQ}
 with the second order $\partial_t^2 R$, 
 so that a delta function impulse gives a linear-ramp 
 instead of a step response.
Then we would no longer use $\partial_t x = v$
 to say that ``velocity causes change in position'', 
 but instead use $\partial_t^2 x = a$
 and say that
 ``acceleration  causes change in position''.
However, 
 %I will not argue here for (or against) requiring this extra smoothness,
 %save to mention the following fact:
 if we demand this, then 
 %our causal responses to be smoother than a step function, 
 we find that the curl Maxwell's equations --
 with their single time derivative --
 no longer count as causal; 
 and other wave equations 
 (e.g. \eqref{eqn-wave-basic}, 
       \eqref{eqn-wave-basic-R}, 
       \eqref{eqn-wave-basic-P})
 suffer a similar fate.

%
% =======================================================================
\section{Summary}\label{S-summary}

KK causal models are constructed 
 as temporal differential equations
 where the changes with time of the effect
 depends on the strength of some cause.
The simplest possible model
 is written down as eqn. \eqref{eqn-basicRQ}, 
 containing a cause $Q$, 
 something to be affected $R$, 
 and a single time derivative applied to $R$.
In such a case 
 an impulsive cause leads to a step-like effect.
For example, 
 in kinematics, 
% the KK causality definition used here along with 
 the equation $\partial_t x = v$, 
 allows us to make the statement that 
 ``velocity $v$ causes a change in position $x$''.
In contrast, 
 writing down e.g. $F=ma$ does not
 allow me to claim either that ``$ma$ causes $F$'' 
 or ``$F/m$ causes $a$''.
But I could instead write $\partial_t v = F/m$ 
 and \emph{then} make the KK causal statement
 ``a given $F/m$ causes a change in velocity $v$''.
This argument then shows that 
Maxwell's equations are unambiguously causal in the KK sense, 
 whether or not any useful alternative definitions of causality exist.

It is well known that a spectral analysis of ``time-step'' causality
 gives rise to the Kramers Kronig relations, 
 which apply constraints to measured spectral data; 
 and link the real and imaginary parts of the spectra
 in a global-to-local manner.
Here I have broadened the notion of ``time-step'' KK causality
 and developed ``wave-cone'' $\Omega$KKK relations
 for signal causality in spacetime,
 a formulation 
 compatible with not only Maxwell's equations,
 but most simple wave equations expressible in a
 second order form.

Lastly, 
 the definition of KK or $\Omega$KKK causality allows us
 to see the validity of Norton's argument \cite{Norton-2009bjps}
 that it is not \emph{a priori} necessary
 to \emph{add} causality as an extra assumption to physical models.
Instead, 
 we should take any given model 
 and determine whether or not it is compatible
 with causality by its construction --
 i.e. from its differential equation(s).

%
% =======================================================================
\bibliography{/home/physics/_work/bibtex.bib}

\begin{thebibliography}{35}
\expandafter\ifx\csname natexlab\endcsname\relax\def\natexlab#1{#1}\fi
\expandafter\ifx\csname bibnamefont\endcsname\relax
  \def\bibnamefont#1{#1}\fi
\expandafter\ifx\csname bibfnamefont\endcsname\relax
  \def\bibfnamefont#1{#1}\fi
\expandafter\ifx\csname citenamefont\endcsname\relax
  \def\citenamefont#1{#1}\fi
\expandafter\ifx\csname url\endcsname\relax
  \def\url#1{\texttt{#1}}\fi
\expandafter\ifx\csname urlprefix\endcsname\relax\def\urlprefix{URL }\fi
\providecommand{\bibinfo}[2]{#2}
\providecommand{\eprint}[2][]{\url{#2}}

\bibitem[{\citenamefont{Lobo}(2008)}]{Lobo-2007}
\bibinfo{author}{\bibfnamefont{F.~S.~N.} \bibnamefont{Lobo}},\\
  \emph{\bibinfo{title}{Nature of time and causality in Physics}}\\
  (\bibinfo{publisher}{Emerald}, \bibinfo{year}{2008}), pp.
  \bibinfo{pages}{395--422}, ISBN \bibinfo{isbn}{978-0-08-046977-5},\\
  %\eprint{arXiv:0710.0428}, \\
  %\urlprefix\url{http://arxiv.org/abs/0710.0428}.
  also \axURL{0710.0428}.

\bibitem[{\citenamefont{Lucas}(1984)}]{Lucas-Causality}
\bibinfo{author}{\bibfnamefont{J.~R.} \bibnamefont{Lucas}},\\
  \emph{\bibinfo{title}{Space, time and causality : an essay in natural
  philosophy}} \\
  (\bibinfo{publisher}{Clarendon}, \bibinfo{address}{Oxford},
  \bibinfo{year}{1984}), ISBN \bibinfo{isbn}{0-19-875058-7}.

\bibitem[{\citenamefont{Pearl}(2000)}]{Pearl-Causality}
\bibinfo{author}{\bibfnamefont{J.}~\bibnamefont{Pearl}},\\
  \emph{\bibinfo{title}{Causality : models, reasoning, and inference}}\\
  (\bibinfo{publisher}{Cambridge University Press}, \bibinfo{year}{2000}), \\
  ISBN \bibinfo{isbn}{978-0-521-77362-1}.

\bibitem[{\citenamefont{Rolnick}(1974)}]{Rolnick-Causality}
\bibinfo{editor}{\bibfnamefont{W.~B.} \bibnamefont{Rolnick}}, ed., \\
  \emph{\bibinfo{title}{Causality and physical theories (Wayne State
  University, 1973)}}, \\
  no.~\bibinfo{number}{16} in \bibinfo{series}{AIP
  Conference Proceedings} (\bibinfo{publisher}{American Institute of Physics},
  \bibinfo{address}{New York}, \bibinfo{year}{1974}), \\
  \XWEB{https://aip.scitation.org/toc/apc/16/1}, \\
  \doiURL{10.1063/1.2948445} (introduction), \\
  ISBN \bibinfo{isbn}{978-0-88318-115-7}.

\bibitem[{\citenamefont{Mermin}(1989)}]{Mermin-1989pt-shutup}
\bibinfo{author}{\bibfnamefont{N.~D.} \bibnamefont{Mermin}},\\
  \bibinfo{journal}{Physics Today} \textbf{\bibinfo{volume}{42}},
  \bibinfo{pages}{9} (\bibinfo{year}{1989}), \\
  %\urlprefix\url{http://dx.doi.org/10.1063/1.2810963}.
  %doi:\href{http://dx.doi.org/10.1063/1.2810963}{10.1063/1.2810963}.
  \doiURL{10.1063/1.2810963}.

\bibitem[{\citenamefont{Bohren}(2010)}]{Bohren-2010ejp}
\bibinfo{author}{\bibfnamefont{C.~F.} \bibnamefont{Bohren}},\\
  \bibinfo{journal}{Eur. J. Phys.} \textbf{\bibinfo{volume}{31}},
  \bibinfo{pages}{573} (\bibinfo{year}{2010}), \\
  %\urlprefix\url{http://iopscience.iop.org/0143-0807/31/3/014}.
  %doi:\href{http://dx.doi.org/10.1088/0143-0807/31/3/014}{10.1088/0143-0807/31/3/014}.
  \doiURL{10.1088/0143-0807/31/3/014}.


\bibitem[{\citenamefont{Landau and Lifshitz}(1984)}]{LandauLifshitz}
\bibinfo{author}{\bibfnamefont{L.~D.} \bibnamefont{Landau}} \bibnamefont{and}
  \bibinfo{author}{\bibfnamefont{E.~M.} \bibnamefont{Lifshitz}},\\
  \emph{\bibinfo{title}{Electrodynamics of Continuous Media}}\\
  (\bibinfo{publisher}{Pergamon}, \bibinfo{address}{Oxford and New York},
  \bibinfo{year}{1984}).


\bibitem[{\citenamefont{Corduneanu}(2002)}]{Corduneanu-FECAUSAL}
\bibinfo{author}{\bibfnamefont{C.} \bibnamefont{Corduneanu}},\\
  \emph{\bibinfo{title}{Functional Equations with Causal Operators}}\\
  (\bibinfo{publisher}{Taylor and Francis}, \bibinfo{address}{London},
  \bibinfo{year}{2002}).

\bibitem[{\citenamefont{Corduneanu}(2000)}]{Corduneanu-2000jde}
\bibinfo{author}{\bibfnamefont{C.}~\bibnamefont{Corduneanu}},
  \\ \bibinfo{journal}{J. Differ. Equations} \textbf{\bibinfo{volume}{168}},
  \bibinfo{pages}{93--101} (\bibinfo{year}{2000}), \\
  \doiURL{10.1006/jdeq.2000.3879}.


\bibitem[{\citenamefont{Kinsler}(2010{\natexlab{a}})}]{Kinsler-2010pra-lfiadc}
\bibinfo{author}{\bibfnamefont{P.}~\bibnamefont{Kinsler}},\\
  \bibinfo{journal}{Phys. Rev. A} \textbf{\bibinfo{volume}{82}},
  \bibinfo{pages}{055804} (\bibinfo{year}{2010}{\natexlab{a}}), \\
  %\eprint{1008.2088},
  %\urlprefix\url{http://link.aps.org/doi/10.1103/PhysRevA.81.013819}.
  \doiURL{10.1103/PhysRevA.81.013819},\\
    also \axURL{1008.2088}.


\bibitem[{\citenamefont{Kinsler}(2017)}]{Kinsler-2014arxiv-negfreq}
  \bibinfo{author}{\bibfnamefont{P.}~\bibnamefont{Kinsler}},\\
  (\bibinfo{year}{2014}), \\
  \bibinfo{note}{``What I talk about when I talk about propagation''}, 
  \\ \XARXIV{1408.0128}

\bibitem[{\citenamefont{Rideout and Sorkin}(2000)}]{Rideout-S-2000prd}
\bibinfo{author}{\bibfnamefont{D.~P.}~\bibnamefont{Rideout}},
\bibinfo{author}{\bibfnamefont{R.~D.}~\bibnamefont{Sorkin}},
  \\ \bibinfo{journal}{Phys. Rev. D} \textbf{\bibinfo{volume}{61}},
  \bibinfo{pages}{024002} (\bibinfo{year}{2000}), \\ \XARXIV{gr-qc/9904062},
  \\ \XDOI{10.1103/PhysRevD.61.024002}.


\bibitem[{\citenamefont{MathWorld}({\natexlab{a}})}]{WfmMathWorld-FourierTrans%
form}
\bibinfo{author}{\bibnamefont{MathWorld}}, \emph{\bibinfo{title}{Fourier
  transform}},\\
  {\scriptsize\url{http://mathworld.wolfram.com/FourierTransform.html}}.

\bibitem[{\citenamefont{Davies}(2002)}]{Davies-IntegralTfs}
\bibinfo{author}{\bibfnamefont{B.}~\bibnamefont{Davies}},\\
  \emph{\bibinfo{title}{Integral Transforms and Their Applications}}\\
  (\bibinfo{publisher}{Springer}, \bibinfo{year}{2002}), ISBN
  \bibinfo{isbn}{978-1-44192-950-1}.

\bibitem[{\citenamefont{Lucarini et~al.}(2005)\citenamefont{Lucarini, Saarinen,
  Peiponen, and Vartiainen}}]{KKinOMR}
\bibinfo{author}{\bibfnamefont{V.}~\bibnamefont{Lucarini}},
  \bibinfo{author}{\bibfnamefont{J.~J.} \bibnamefont{Saarinen}},
  \bibinfo{author}{\bibfnamefont{K.~E.} \bibnamefont{Peiponen}},
  %\bibnamefont{and} 
  \bibinfo{author}{\bibfnamefont{E.~M.}
  \bibnamefont{Vartiainen}}, \\
  \emph{\bibinfo{title}{Kramers-Kronig Relations in Optical Materials Research}}, \\
  vol. \bibinfo{volume}{110}
  (\bibinfo{publisher}{Springer}, \bibinfo{address}{Berlin / Heidelberg},
  \bibinfo{year}{2005}), ISBN \bibinfo{isbn}{978-3-540-23673-3},
  %\urlprefix\url{http://www.springerlink.com/content/p103l4x87026/?p=8033dc2dd%12a42edb7e43c160dfdf813&pi=2}.
  \doiURL{10.1007/b138913}.

\bibitem[{\citenamefont{Toll}(1956)}]{Toll-1956pr}
\bibinfo{author}{\bibfnamefont{J.~S.} \bibnamefont{Toll}},\\
  \bibinfo{journal}{Phys. Rev.} \textbf{\bibinfo{volume}{104}},
  \bibinfo{pages}{1760} (\bibinfo{year}{1956}), \\
  \doiURL{10.1103/PhysRev.104.1760}.

\bibitem[{\citenamefont{MathWorld}({\natexlab{b}})}]{WfmMathWorld-HilbertTrans%
form}
\bibinfo{author}{\bibnamefont{MathWorld}}, \emph{\bibinfo{title}{Hilbert
  transform}},\\
  %\urlprefix
  {\scriptsize\url{http://mathworld.wolfram.com/HilbertTransform.html}}.

\bibitem[{\citenamefont{Titchmarsh}(1967)}]{Titchmarsh-FI}
\bibinfo{author}{\bibfnamefont{E.~C.} \bibnamefont{Titchmarsh}},\\
  \emph{\bibinfo{title}{Introduction to the theory of Fourier Integrals}}\\
  (\bibinfo{publisher}{Clarendon Press}, \bibinfo{address}{Oxford},
  \bibinfo{year}{1967}), \bibinfo{edition}{corrected 2nd} ed.

\bibitem[{\citenamefont{MathWorld}({\natexlab{c}})}]{WfmMathWorld-PrincipalVal%
ue}
\bibinfo{author}{\bibnamefont{MathWorld}}, \emph{\bibinfo{title}{Cauchy
  principal value}},\\
  %\urlprefix
  {\scriptsize\url{http://mathworld.wolfram.com/CauchyPrincipalValue.html}}.

\bibitem[{\citenamefont{Kinsler}(2009)}]{Kinsler-2009pra}
\bibinfo{author}{\bibfnamefont{P.}~\bibnamefont{Kinsler}},\\
  \bibinfo{journal}{Phys. Rev. A} \textbf{\bibinfo{volume}{79}},
  \bibinfo{pages}{023839} (\bibinfo{year}{2009}), \\
  %\eprint{0901.2466},
  %\urlprefix\url{http://link.aps.org/abstract/PRA/v79/e023839}.
  \doiURL{10.1103/PhysRevA.79.023839},\\
    also \axURL{0901.2466}.

\bibitem[{\citenamefont{Stockman}(2007)}]{Stockman-2007prl}
\bibinfo{author}{\bibfnamefont{M.~I.} \bibnamefont{Stockman}},\\
  \bibinfo{journal}{Phys. Rev. Lett.} \textbf{\bibinfo{volume}{98}},
  \bibinfo{pages}{177404} (\bibinfo{year}{2007}),\\
  %\urlprefix\url{http://link.aps.org/abstract/PRL/v98/e177404}.
  %doi:\href{http://dx.doi.org/10.1103/PhysRevLett.98.177404}{10.1103/PhysRevLett.98.177404}.
  \doiURL{10.1103/PhysRevLett.98.177404}


\bibitem[{\citenamefont{Kinsler and McCall}(2008)}]{Kinsler-M-2008prl}
\bibinfo{author}{\bibfnamefont{P.}~\bibnamefont{Kinsler}} \bibnamefont{and}
  \bibinfo{author}{\bibfnamefont{M.~W.} \bibnamefont{McCall}},\\
  \bibinfo{journal}{Phys. Rev. Lett.} \textbf{\bibinfo{volume}{101}},
  \bibinfo{pages}{167401} (\bibinfo{year}{2008}), \\
  %\eprint{arXiv:0812.1521},
  %\urlprefix\url{http://link.aps.org/abstract/PRL/v101/e167401}.
  %doi:\href{http://dx.doi.org/10.1103/PhysRevLett.101.167401}{10.1103/PhysRevLett.101.167401}.
  \doiURL{10.1103/PhysRevLett.101.167401},\\
  also \axURL{0812.1521}.


\bibitem[{\citenamefont{MathWorld}({\natexlab{d}})}]{WfmMathWorld-DampedSHM}
\bibinfo{author}{\bibnamefont{MathWorld}}, \emph{\bibinfo{title}{Damped simple
  harmonic motion}}, \\
  {\scriptsize\url{http://mathworld.wolfram.com/DampedSimpleHarmonicMotion.html}}.

\bibitem[{\citenamefont{Wikipedia}()}]{Wikipedia-DampedSHM}
\bibinfo{author}{\bibnamefont{Wikipedia}}, \emph{\bibinfo{title}{Harmonic
  oscillator}}, \bibinfo{note}{(as at 20 September 2011)}, 
  {\scriptsize\url{http://en.wikipedia.org/wiki/Harmonic_oscillator}}.

\bibitem[{\citenamefont{Reitz et~al.}(1980)\citenamefont{Reitz, Milford, and
  Christy}}]{RMC}
\bibinfo{author}{\bibfnamefont{J.~R.} \bibnamefont{Reitz}},
  \bibinfo{author}{\bibfnamefont{F.~J.} \bibnamefont{Milford}},
  \bibnamefont{and} \bibinfo{author}{\bibfnamefont{R.~W.}
  \bibnamefont{Christy}}, \\
  \emph{\bibinfo{title}{Foundations of electromagnetic theory}} \\
  (\bibinfo{publisher}{Addison-Wesley}, \bibinfo{year}{1980}),
  \bibinfo{edition}{3rd} ed.

\bibitem[{\citenamefont{Jackson}(1975)}]{Jackson-ClassicalED}
\bibinfo{author}{\bibfnamefont{J.~D.} \bibnamefont{Jackson}},\\
  \emph{\bibinfo{title}{Classical Electrodynamics}}\\
  (\bibinfo{publisher}{Wiley}, \bibinfo{year}{1975}), ISBN
  \bibinfo{isbn}{0-471-43132-X}.

\bibitem[{\citenamefont{Pendry et~al.}(1999)\citenamefont{Pendry, Holden,
  Robbins, and Stewart}}]{Pendry-HRS-1999ieeemtt}
\bibinfo{author}{\bibfnamefont{J.~B.} \bibnamefont{Pendry}},
  \bibinfo{author}{\bibfnamefont{A.~J.} \bibnamefont{Holden}},
  \bibinfo{author}{\bibfnamefont{D.~J.} \bibnamefont{Robbins}},
  \bibnamefont{and} \bibinfo{author}{\bibfnamefont{W.~J.}
  \bibnamefont{Stewart}}, \\
  \bibinfo{journal}{IEEE Transactions on Microwave
  Theory and Techniques} \textbf{\bibinfo{volume}{47}}, \bibinfo{pages}{2075}
  (\bibinfo{year}{1999}), \\
  {\scriptsize\url{http://ieeexplore.ieee.org/search/wrapper.jsp?arnumber=798002}}.

\bibitem[{\citenamefont{Hehl and Obukhov}(2003)}]{HehlObukhov}
\bibinfo{author}{\bibfnamefont{F.~W.} \bibnamefont{Hehl}} \bibnamefont{and}
  \bibinfo{author}{\bibfnamefont{Y.~N.} \bibnamefont{Obukhov}},\\
  \bibinfo{journal}{Progress in Mathematical Physics}
  \textbf{\bibinfo{volume}{22}} (\bibinfo{year}{2003}).

\bibitem[{\citenamefont{Savage}(2011)}]{Savage-2011arXiv-cem}
\bibinfo{author}{\bibfnamefont{C.~M.} \bibnamefont{Savage}}\\
  \bibinfo{journal}{The Physics Teacher}
  \textbf{\bibinfo{volume}{50}},
  \bibinfo{pages}{226} (\bibinfo{year}{2012}), \\
  also \axURL{1105.1197}.

\bibitem[{\citenamefont{Jefimenko}(2004)}]{Jefimenko-2004ejp}
\bibinfo{author}{\bibfnamefont{O.}~\bibnamefont{Jefimenko}},\\
  \bibinfo{journal}{Eur. J. Phys.} pp. \bibinfo{pages}{287--296}
  (\bibinfo{year}{2004}),\\
  %\urlprefix\url{http://iopscience.iop.org/0143-0807/25/2/015}.
  %doi:\href{http://dx.doi.org/10.1088/0143-0807/25/2/015}{10.1088/0143-0807/25/2/015}.
  \doiURL{10.1088/0143-0807/25/2/015}.

\bibitem[{\citenamefont{Taflove}(1995)}]{Taflove-FDTD}
\bibinfo{author}{\bibfnamefont{A.}~\bibnamefont{Taflove}},\\
  \emph{\bibinfo{title}{Computational Electrodynamics: The Finite-Difference
  Time-Domain Method}} \\
  (\bibinfo{publisher}{Artech},
  \bibinfo{address}{Norwood}, \bibinfo{year}{1995}).

\bibitem[{\citenamefont{Heras}(2011)}]{Heras-2011ajp}
\bibinfo{author}{\bibfnamefont{J.~A.} \bibnamefont{Heras}},\\
  \bibinfo{journal}{Am. J. Phys.} \textbf{\bibinfo{volume}{79}},
  \bibinfo{pages}{409} (\bibinfo{year}{2011}), \\
  %\urlprefix\url{http://link.aip.org/link/ajpias/v79/i4/p409/s1}.
  %doi:\href{http://dx.doi.org/10.1119/1.3533223}{10.1119/1.3533223}.
  \doiURL{10.1119/1.3533223}.

\bibitem[{\citenamefont{Kinsler}(2017)}]{Kinsler-2017arxiv-faradin}
  \bibinfo{author}{\bibfnamefont{P.}~\bibnamefont{Kinsler}},\\
  (\bibinfo{year}{2017}), \\
  \bibinfo{note}{``Faraday's Law and Magnetic Induction: cause and effect''}, 
  \\ \XARXIV{1705.08406}

\bibitem[{\citenamefont{Morse and Ingard}(1968)}]{MorseIngard-Acoustics}
\bibinfo{author}{\bibfnamefont{P.~M.} \bibnamefont{Morse}} \bibnamefont{and}
  \bibinfo{author}{\bibfnamefont{K.~U.} \bibnamefont{Ingard}},\\
  \emph{\bibinfo{title}{Theoretical Acoustics}}\\
  (\bibinfo{publisher}{McGraw-Hill}, \bibinfo{address}{New York},
  \bibinfo{year}{1968}).

\bibitem[{\citenamefont{Kinsler}(2010{\natexlab{b}})}]{Kinsler-2010pra-fchhg}
\bibinfo{author}{\bibfnamefont{P.}~\bibnamefont{Kinsler}},\\
  \bibinfo{journal}{Phys. Rev. A} \textbf{\bibinfo{volume}{81}},
  \bibinfo{pages}{013819} (\bibinfo{year}{2010}{\natexlab{b}}), \\
  %\urlprefix\url{http://link.aps.org/doi/10.1103/PhysRevA.81.013819}.
  %doi:\href{http://dx.doi.org/10.1103/PhysRevA.81.013819}{10.1103/PhysRevA.81.013819}.
  \doiURL{10.1103/PhysRevA.81.013819}, \\
  also \axURL{0810.5689}.

\bibitem[{\citenamefont{Kinsler}(2010{\natexlab{c}})}]{Kinsler-2010pra-dblnlGpm}
\bibinfo{author}{\bibfnamefont{P.}~\bibnamefont{Kinsler}},\\
  \bibinfo{journal}{Phys. Rev. A} \textbf{\bibinfo{volume}{81}},
  \bibinfo{pages}{023808} (\bibinfo{year}{2010}{\natexlab{c}}), \\
  %\eprint{0909.3407},
  %\urlprefix\url{http://pra.aps.org/abstract/PRA/v81/i2/e023808}.
  \doiURL{10.1103/PhysRevA.81.023808},\\ 
  also \axURL{0909.3407}.

\bibitem[{\citenamefont{Todd and Menicucci}(2017)}]{Todd-M-2017fp}
  \bibinfo{author}{\bibfnamefont{S.~L.}~\bibnamefont{Todd}},
  \bibinfo{author}{\bibfnamefont{N.~C.}~\bibnamefont{Menicucci}},\\
  \bibinfo{journal}{Found. Phys.} \textbf{\bibinfo{volume}{47}},
  \bibinfo{pages}{1267--1293} (\bibinfo{year}{2017}, \\
  \doiURL{10.1007/s10701-017-0109-0}.

\bibitem[{\citenamefont{Schutz}(1986)}]{Schutz-FCRelativity}
\bibinfo{author}{\bibfnamefont{B.~F.} \bibnamefont{Schutz}},\\
  \emph{\bibinfo{title}{A first course in general relativity}}\\
  (\bibinfo{publisher}{Cambridge University Press}, \bibinfo{year}{1986}), \\
  ISBN \bibinfo{isbn}{0-521-27703-5}.

\bibitem[{\citenamefont{MathWorld}({\natexlab{e}})}]{WfmMathWorld-MellinTransf%
orm}
\bibinfo{author}{\bibnamefont{MathWorld}}, 
  \emph{\bibinfo{title}{Mellin transform}},\\
  {\scriptsize\url{http://mathworld.wolfram.com/MellinTransform.html}}.

\bibitem[{\citenamefont{Weinstein}(2011)}]{Weinstein-2011mpla}
\bibinfo{author}{\bibfnamefont{S.}~\bibnamefont{Weinstein}},\\
  \bibinfo{journal}{Mod. Phys. Lett. A} \textbf{\bibinfo{volume}{26}},
  \bibinfo{pages}{815} (\bibinfo{year}{2011}), \\
  %\eprint{1004.1346},
  {\scriptsize\url{http://linkinghub.elsevier.com/retrieve/pii/S0217732311035298}},
  also \axURL{1004.1346}.

\bibitem[{\citenamefont{Norton}(2009)}]{Norton-2009bjps}
\bibinfo{author}{\bibfnamefont{J.~D.} \bibnamefont{Norton}},\\
  \bibinfo{journal}{Brit. J. Phil. Sci.} \textbf{\bibinfo{volume}{60}},
  \bibinfo{pages}{475} (\bibinfo{year}{2009}),\\
  %\urlprefix\url{http://bjps.oxfordjournals.org/content/60/3/475}.
  %doi:\href{http://dx.doi.org/10.1093/bjps/axp030}{10.1093/bjps/axp030}.
  \doiURL{10.1093/bjps/axp030}

\end{thebibliography}

%
% =======================================================================
\section*{Appendix A: A discrete picture}\label{S-appdx-discrete}

\begin{figure}
\includegraphics[angle=-0,width=0.82\columnwidth]{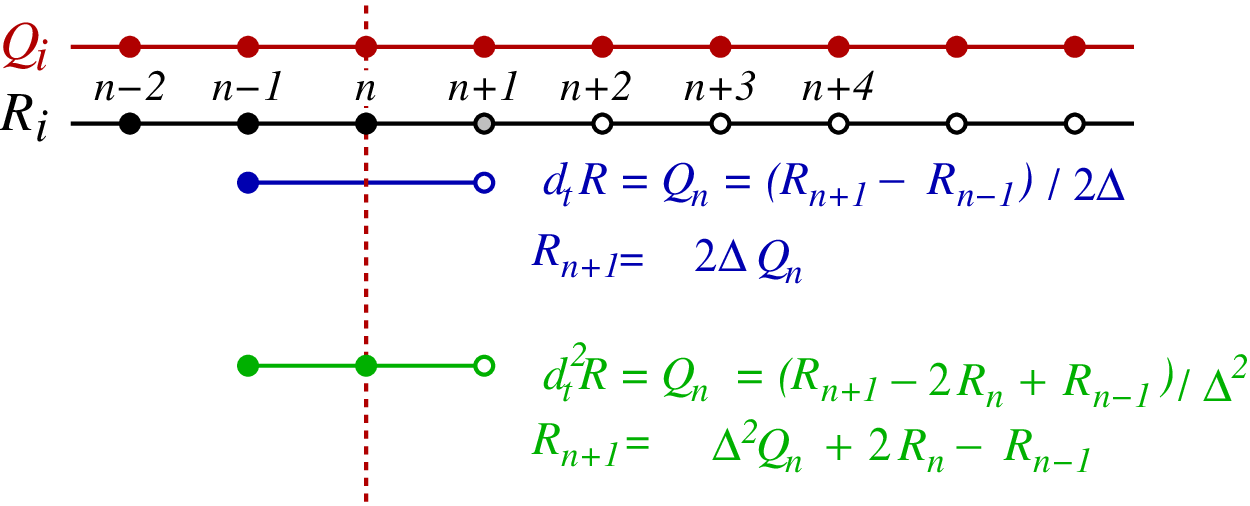}
\caption{
Here the filled circles denote known information, 
 while unknown information is denoted by open circles.
All of the causes $Q_i$ on the top line are specified by the model, 
 and so are ``known''.
Knowledge of $R_i$, 
 indicated on the second line, 
 has to be computed, 
 and so at some index $i=n$ we only know $R_i$ up to $R_n$, 
 and not $R_{n+1}$ and later.
The lower parts 
 (blue, then green at the bottom)
 compare two simple causal models, 
 as discussed in the text.
The $R_{n+1}$ circle is shaded in grey
 to denote the fact that it is the quantity that 
 will be next computed:
 the change from $R_n$ to $R_{n+1}$ is the ``effect''.
}
\label{fig-discrete-dtA}
\end{figure}

We can also motivate the basic causal prescription of 
 section \ref{S-causalDE}
 using a discretized argument
 of the type used to numerically solve differential equations. 
This is introduced by the diagram in fig. \ref{fig-discrete-dtA}.

For the simple causal model $\partial_t R = Q$
 defining an effect on $R$ due to the cause $Q$, 
 the information needed to calculate the next value of $R_i$, 
 i.e. $R_{n+1}$, 
 consists of the known $Q_n$
 and the past (known) value of $R_{n-1}$.
For an alternative causal model $\partial_t^2 R = Q$, 
 the information needed to calculate $R_{n+1}$, 
 consists of the known values $Q_n$, $R_n$ and $R_{n-1}$.
Again, 
 this demonstrates that we can calculate the future ($R_{n+1}$)
 using only known -- current or past -- 
 information.

It is always the highest order discrete derivative of $R$
 in the model which reaches furthest back and forwards in time
 when being calculated; 
 and it is the most future-like component which we need
 to calculate, 
 but \emph{only} on the basis of known information.
We can see on fig. \ref{fig-discrete-dtA}
 how to do the necessary rearrangements for the two simple models; 
 but the general procedure is to discretize the 
 entire differential equation around
 some central time-index $i=n$ and then rearrange
 so that the most future-like term is on the LHS
 with the rest on the RHS.
In the case of a third or fourth order derivatives centered on $i=n$, 
 the most future term will be $R_{n+2}$, 
 which might seem problematic.
However, 
 the fix is simple -- 
 we just re-centre on $n-1$ so that 
 the most future term will instead become the desired $R_{n+1}$.
Happily, 
 we can apply this re-centering as required using even larger shifts,
 if still higher order derivatives are present.
As a result we can always explicitly calculate 
 the future ($R_{n+1}$) using only known information
 from $i \le n$.

We might speculate that there could perhaps be a way 
 to invert this scheme and instead 
 denote $\partial_t R$ as the cause and $Q$ the effect.
However, 
 even for the simple models, 
 we see that to calculate $Q_n$ we need $R_{n+1}$.
Although all of $R_i$ might indeed be pre-specified by our model, 
 it is nevertheless true that when evaluating $Q_n$
 we should not (yet) have access to $R_{n+1}$ --
 that value of ``cause'' has not been expressed to $R$ yet.
Thus we cannot consider $R$ as the cause,
 because we insist that only
 current or past information \emph{known} to what is being affected
 can influence it.

%
% =======================================================================
\section*{Appendix B: Types of cause}\label{S-appdx-types}

As already discussed in section \ref{S-simul}, 
 there is a school of thought in EM 
 that would rather that currents and charges
 were given a primary status as causes, 
 and is opposed to fields being treated as causes as a matter of principle.

Perhaps the most compelling basis for this position
 is that any spacetime (i.e. \emph{moving}) transformation 
 alters the definitions of the space and time derivatives; 
 thus even the causal Maxwells equations
 of eqn. \eqref{eqn-vaccumcurlMaxwell}
 are changed in a moving frame\footnote{Although suitable redefinition
  of the fields preserves the original structure.}.
Specifically, 
 in transformation to a moving frame, 
 a ``cause'' term originally based solely on a spatial derivative
 can gain a time derivative component -- 
 i.e. part of a cause has been re-expressed 
 in the new frame as an effect.
Likewise an effect (time derivative) term can be transformed into one
 with a spatial derivative -- 
 or ``cause-like'' part.
This means that 
 observers in different frames can attribute
 cause and effect differently, 
 i.e. the attribution is not \emph{frame independent}.

In a practical sense, 
 the ``frame-dependence'' criticism can be addressed simply
 by dividing causes into two types:

\begin{description}

\item[Primary] causes are those that %[{Type 1}]
 appear without any spatial or temporal derivatives,
 and include source terms such as currents $\Vec{J} $ and charges $\rho$ in EM.
Regardless of frame of reference, 
 where they might be expressed differently,
 according to the scheme outlined in this paper,
 these are always considered to be causes.

\item[Secondary] causes 
 appear as (with) derivative terms, 
 and include terms such 
 the $\grad \cross \Vec{E}$ and $\grad \cross \Vec{H}$ in EM.
These are frame dependent, 
 and although we would usually expect such a term to persist, 
 regardless of frame, 
 its particular expression or values can and will differ.

\end{description}

The case of two observers with different 4-velocities 
 crossing at a point is an important one, 
 because it exhibits the discrepancy in attribution of cause without 
 any complicating factors from having separate locations. 
Their past lightcones coincide, 
 but because their rest frames differ --
 and assuming their rest frames are their preferred frames --
 the observers will likewise prefer
 different attributions of cause and effect\footnote{However, 
  such observers could still chose to agree
  on some preferred frame,
  e.g. the rest frame of a nearby mass, or of surrounding matter.
  In such a case they would also agree on attributions of cause and effect.}.
Although this may trouble the purist, 
 it should not be that surprising, 
 and objections can be mollified by noting that:

\begin{enumerate}

\item
 Since distinct observers, 
 following different trajectories can disagree
 about the ordering of spacetime events, 
 it is hardly unsurprising that they might also 
 disagree about their particular experience of (local) causality.

\item
 Even though they might have distinct perspectives, 
 different observers can still reconcile their respective viewpoints
 with one another.

\end{enumerate}

The scheme followed in the main part of this paper
 provides a practical,
 straightforward, 
 and local way to systematically attribute cause and effect.
However, 
 those interested in a more abstract or philosophical approache, 
 or those reconciling the views of disparate observers,
 should distinguish between primary and secondary causes.
%Note that insisting that those terms labelled ``secondary causes'' 
% should not be regarded as any sort of cause at all
% will come into conflict with the simpler, 
% practical view naturally taken by a sole observer.

%
% =======================================================================
\section*{Appendix C: Models with ambiguous predictions}\label{S-appdx-exotics}

\def\pPderivT{\partial_t}

As pointed out to me by Jonathan Gratus (JG), 
 the causal interpretation offered in this paper 
 works best with mathematical models that can be decomposed
 into first order linear or quasi-linear temporal differential equations.
As an example of the complications that can arise, 
 JG pointed out that a model might follow an equation
 such as the following: 
~
\begin{align}
  \left(
    \pPderivT R
  \right)^2
&=
  R
,
\label{eqn-dtr2-squared}
\\
\textrm{or} \qquad
    \pPderivT R
&=
  %\textup{sgn}(R)
 \pm
  \sqrt{R}
.
\label{eqn-dtr2-sqroot}
\end{align}
 where to make the discussion simpler 
 we stipulate that the preferred sign of the square root 
 ensures that $R \ge 0$.
This preference merely reduces the number
 of possible cases we need to consider, 
 and does not change the discussion of causal interpretation.
% is the same as for $R$ itself.
If solved mathematically, 
 we find that the possible solutions can be constructed from 
 pieces where either 
~
\begin{align}
  R(t)&=0
\label{eqn-Rzero}
\\
 \textrm{or}\qquad
 R(t)&=\left(t-t_n\right)^2/4, \quad \textrm{for any}~~ t_n \in \mathbb{R}.
\label{eqn-Rparabola}
\end{align}

\begin{figure}
\includegraphics[angle=-0,width=0.98\columnwidth]{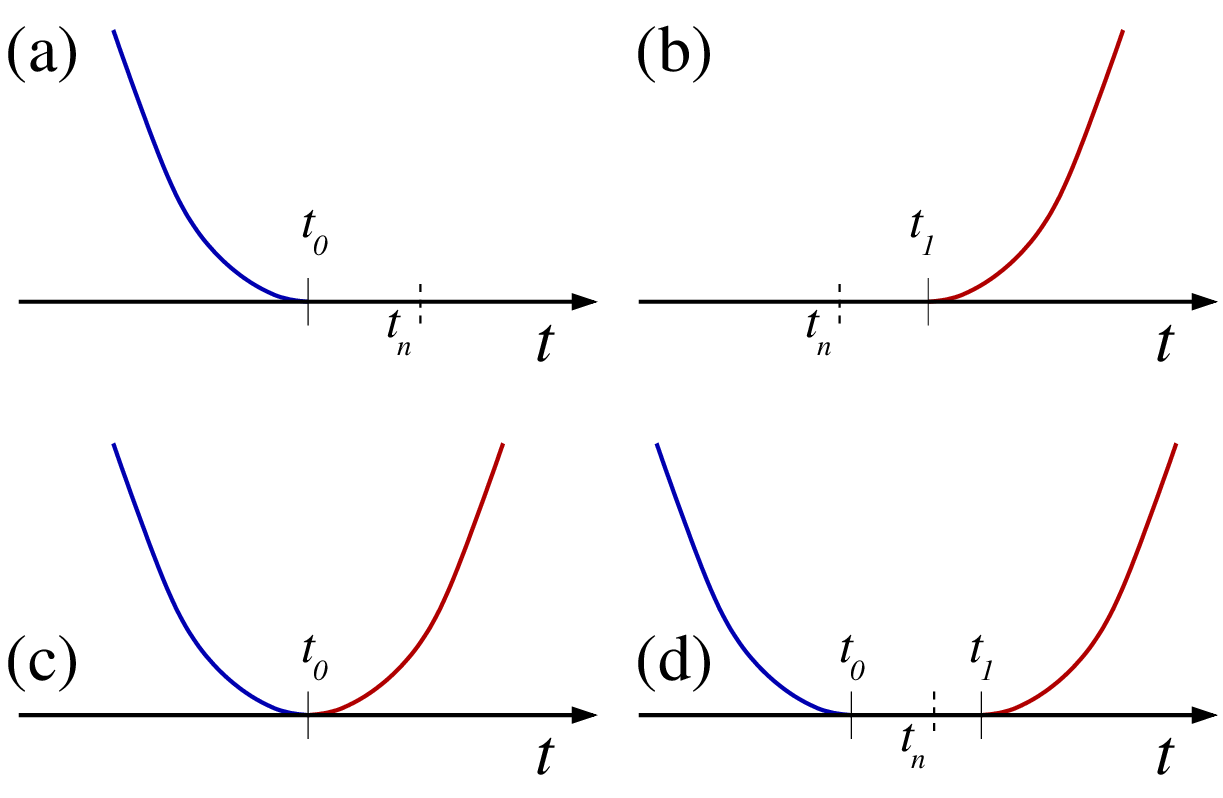}
\caption{
Some possible solutions for the all-times behavior of $R(t)$.
% with the exception of the uninteresting $R(t)=0$ case.
If $R$ is initially non-zero and increasing as in (a,c,d), 
 it eventually reaches $R=0$.
It might then stay zero as in (a), 
 stay parabolic but start increasing again as in (c), 
 or stay zero for a time $t_1-t_0$ before
 resuming its parabolic behaviour (d).
Even if $R$ started at zero, 
 it might start to increase parabolically as in (b). 
}
\label{fig-unusual}
\end{figure}

The main issue of relevance to our causal interpretation 
 is that for any $t_n$ where $R(t_n)=0$,
 $R(t)$ may subsequently either remain at $0$ or become parabolic;
 there is no way to know, 
 based on current or past information,
 which ``prediction'' of this model to prefer.
This is true even if $R(t)=0$ on some finite interval prior to $t_n$, 
 as in panels (a,b,d) of fig. \ref{fig-unusual}. %, 
% where either (or both of) $t_0 < t_n$ and $t_n < t_1$ hold.
As we integrate forward in time to follow the effects on $R$
 of the value of its square root $\sqrt{R}$, 
 at every point where $R=0$
 we have no way to choose which ``effect'' to pick.

Notably,  
 the different choices of possible effect
 differ in their second time derivative, 
 i.e. $\pPderivT^2 R$, 
 a quantity not specified by the model.
There are four possible second derivatives $d_2$ when $R=0$: 
 (a) $d_2=0$,
  for when $R=0$ and remains so; 
 (b) $d_2=1$, 
  when $R$ follows the parabolic solution of \eqref{eqn-Rparabola}; 
 and 
 (c) $d_2$ undefined, 
  when $R$ switches from the $R=0$ solution to the parabolic one
  (or vice versa).

There are two ways to consider mathematical models such as these, 
 where our causal interpretation does not uniquely specify
 the outcome of our named ``effect''
 based on our named ``causes''.

\begin{description}

\item[(i)]
 We might maintain that \eqref{eqn-dtr2-sqroot}
 \emph{does} have the same causal interpretation as proposed here --
 i.e. that changes in $R$ are caused by the value of $\sqrt{R}$, 
 but that the model is insufficiently well specified 
 to provide useful predictions if $R=0$.

\item[(ii)]
 We might say that {because} the model in  \eqref{eqn-dtr2-sqroot}
 is insufficiently well specified, 
 and so does not (or cannot) provide useful predictions, 
 we \emph{should not} make or claim 
 any casual interpretation.

\end{description}

It is of course very reasonable
 to take position (ii), 
 and demand that where that any casual interpretation --
 including the one used in this paper --
 should be explicitly scoped so as to exclude such cases
 where the predictions of a model might be ambiguous.

However,
 I propose here that we should prefer the position (i),
 because the difficulty is the result of the model, 
 not the interpretation; 
 the difficulty in getting a solution exists
 \emph{even in the absence of any attempt at a causal interpretation}.
Further, 
 this choice also means that the basic rules for attributing causality
 remain simple, 
 even though this example shows that those rules 
 are not a guarantee that predicting the nature of effects is straightforward.

\noindent\emph{Postscript:}
As an interesting aside,
 \emph{if}  we know that at some {future} time $t_f$
 that $R(t_f)$ has a non-zero value, 
 then we can at least calculate the time $t_1$
 when $R$ started to move away from zero
 on its parabolic trajectory --
 so that forward-looking ambiguity is resolved.
However,  
 this would not be a causal prediction
 about the future based on  current or past information; 
 it is instead a \emph{retrodiction}
 about the past,
 made using information only available at time $t_f$.
Further, 
 if we try to retrodict further back into the past 
 (i.e. for some $t<t_1$), 
 then the same ambiguity returns:
 did $R$ stay parabolic, or settle at $R=0$, 
 and if it did settle, for what interval\footnote{Anyone who
  at this point 
  decides to introduce proposed knowledge about both 
  the initial conditions $R(t_i)$ at $t_i$
  and the final state $R(t_f)$ at $t_f$
  should probably excuse themselves from offering opinions
  regarding possible causal interpretation --
  because they clearly prefer cases where such is unnecessary!}?

%
% =======================================================================
\clearpage

%%
%% =======================================================================
%\bibliography{/home/physics/_work/bibtex.bib}

%
% =======================================================================
% =======================================================================
\end{document}